\documentclass[preprint]{aastex63}
\newcommand{\ms}{$\mathrm{m\,s^{-1}}$}
\newcommand{\kms}{$\mathrm{km\,s^{-1}}$}
\newcommand\vsini{$v$\,sin\,$i_\star$}

\received{June 1, 2019}
\revised{January 10, 2019}
\accepted{\today}
\submitjournal{AJ}

\shorttitle{The $\pi$ Men Planetary System}
\shortauthors{Hatzes et al.}
\begin{document}

\title{A Radial Velocity Study of the Planetary System of $\pi$ Mensae:\\ Improved Planet Parameters for $\pi$ Mensae\,c and a Third Planet on a 125-d Orbit} 
%\footnote{Released on June, 10th, 2019}}

\correspondingauthor{Artie Hatzes} \email{artie@tls-tautenburg.de} 

\author[0000-0002-3404-8358]{Artie P. Hatzes}
\affiliation{Th\"uringer Landessternwarte, Sternwarte 5, D-07778 Tautenburg, Germany}
\author[0000-0001-8627-9628]{Davide Gandolfi}
\affiliation{Dipartimento di Fisica, Universit\`a degli Studi di Torino, via Pietro Giuria 1, I-10125, Torino, Italy}
\author[0000-0002-0076-6239]{Judith Korth}
\affiliation{Department of Space, Earth and Environment, Astronomy and Plasma Physics, Chalmers University of Technology, SE-412 96 Gothenburg, Sweden}
\author[0000-0003-0650-5723]{Florian Rodler}
\affiliation{European Southern Observatory (ESO), Alonso de C\'ordova 3107, Vitacura, Casilla 19001, Santiago de Chile}
\author[0000-0001-9078-5574]{Silvia Sabotta}
\affiliation{Landessternwarte, Zentrum f\"ur Astronomie der Universit\"at Heidelberg,
K\"onigstuhl 12, 69117 Heidelberg, Germany}
\author{Massimiliano Esposito}
\affiliation{Th\"uringer Landessternwarte, Sternwarte 5, D-07778 Tautenburg, Germany}
\author[0000-0003-0563-0493]{Oscar Barrag\'an}
\affiliation{Sub-department of Astrophysics, Department of Physics, University of Oxford, Oxford, OX1 3RH, UK}
\author[0000-0001-5542-8870]{Vincent Van Eylen}
\affiliation{Mullard Space Science Laboratory, University College London, Holmbury St Mary, Dorking, Surrey, RH5 6NT, UK}
\author[0000-0002-4881-3620]{John~H.~Livingston}
\affiliation{Department of Astronomy, University of Tokyo, 7-3-1 Hongo, Bunkyo-ku, Tokyo 113-0033, Japan}
\author[0000-0001-9211-3691]{Luisa Maria Serrano}
\affiliation{Dipartimento di Fisica, Universit\`a degli Studi di Torino, via Pietro Giuria 1, I-10125, Torino, Italy}
\author[0000-0002-4671-2957]{Rafael Luque}
\affiliation{Instituto de Astrof\'isica de Andaluc\'ia (IAA-CSIC), E-18008, Granada, Spain}
\author{Alexis M.\,S. Smith}
\affiliation{Institute of Planetary Research, German Aerospace Center, Rutherfordstrasse 2, 12489 Berlin, Germany}
\author[0000-0003-3786-3486]{Seth Redfield}
\affiliation{Astronomy Department and Van Vleck Observatory, Wesleyan University, Middletown, CT 06459, USA}
\author[0000-0003-1257-5146]{Carina~M.~Persson}
\affiliation{Department of Space, Earth and Environment, Chalmers University of Technology, Onsala Space Observatory, 439 92 Onsala, Sweden}
\author{Martin P\"atzold}
\affiliation{Rheinisches Institut f\"ur Umweltforschung an der Universit\"at zu K\"oln, Aachener Strasse 209, 50931 K\"oln, Germany}
\author[0000-0003-0987-1593]{Enric Palle}
\affiliation{Instituto de Astrof\'\i sica de Canarias, C/\,V\'\i a L\'actea s/n, 38205 La Laguna, Spain}
\affiliation{Departamento de Astrof\'isica, Universidad de La Laguna, 38206 La Laguna, Spain}
\author[0000-0002-7031-7754]{Grzegorz Nowak}
\affiliation{Instituto de Astrof\'\i sica de Canarias, C/\,V\'\i a L\'actea s/n, 38205 La Laguna, Spain}
\affiliation{Departamento de Astrof\'isica, Universidad de La Laguna, 38206 La Laguna, Spain}
\author{Hannah L. M. Osborne}
\affiliation{Mullard Space Science Laboratory, University College London, Holmbury St Mary, Dorking, Surrey, RH5 6NT, UK}
\author{Norio Narita}
\affiliation{Department of Astronomy, University of Tokyo, 7-3-1 Hongo, Bunkyo-ku, Tokyo 113-0033, Japan}
\affiliation{Instituto de Astrof\'\i sica de Canarias, C/\,V\'\i a L\'actea s/n, 38205 La Laguna, Spain}
\affiliation{Astrobiology Center, NINS, 2-21-1 Osawa, Mitaka, Tokyo 181-8588, Japan}
\affiliation{National Astronomical Observatory of Japan, NINS, 2-21-1 Osawa, Mitaka, Tokyo 181-8588, Japan}
\affiliation{JST, PRESTO, 7-3-1 Hongo, Bunkyo-ku, Tokyo 113-0033, Japan}
\author[0000-0002-0129-0316]{Savita Mathur}
\affiliation{Instituto de Astrof\'\i sica de Canarias, C/\,V\'\i a L\'actea s/n, 38205 La Laguna, Spain}
\affiliation{Departamento de Astrof\'isica, Universidad de La Laguna, 38206 La Laguna, Spain}
\author{Kristine~W.\,F.~Lam}
\affiliation{Center for Astronomy and Astrophysics, TU Berlin, Hardenbergstr. 36, 10623 Berlin, Germany}
\author[0000-0002-1623-5352]{Petr Kab\'ath}
\affiliation{Astronomical Institute, Czech Academy of Sciences, Fri\v{c}ova 298, 25165, Ond\v{r}ejov, Czech Republic}
\author{Marshall C. Johnson}
\affiliation{Las Cumbres Observatory, 6740 Cortona Dr., Ste. 102, Goleta, CA 93117, USA}
\author{Eike W. Guenther}
\affiliation{Th\"uringer Landessternwarte Tautenburg, Sternwarte 5, D-07778 Tautenberg, Germany}
\author{Sascha Grziwa}
\affiliation{Rheinisches Institut f\"ur Umweltforschung an der Universit\"at zu K\"oln, Aachener Strasse 209, 50931 K\"oln, Germany}
\author{Elisa Goffo}
\affiliation{Th\"uringer Landessternwarte, Sternwarte 5, D-07778 Tautenburg, Germany}
\affiliation{Dipartimento di Fisica, Universit\`a degli Studi di Torino, via Pietro Giuria 1, I-10125, Torino, Italy}
\author[0000-0002-0855-8426]{Malcolm Fridlund}
\affiliation{Department of Space, Earth and Environment, Chalmers University of Technology, Onsala Space Observatory, 439 92 Onsala, Sweden}
\affiliation{Leiden Observatory, University of Leiden, PO Box 9513, 2300 RA, Leiden, The Netherlands\label{Leiden}}
\author{Michael Endl}
\affiliation{Department of Astronomy and McDonald Observatory, University of Texas at Austin, 2515 Speedway,~Stop~C1400,~Austin,~TX~78712,~USA}
\author[0000-0003-0047-4241]{Hans J. Deeg}
\affiliation{Instituto de Astrof\'\i sica de Canarias, C/\,V\'\i a L\'actea s/n, 38205 La Laguna, Spain}
\affiliation{Departamento de Astrof\'isica, Universidad de La Laguna, 38206 La Laguna, Spain}
\author[0000-0001-6803-9698]{Szilard Csizmadia}
\affiliation{Institute of Planetary Research, German Aerospace Center, Rutherfordstrasse 2, 12489 Berlin, Germany}
\author{William D. Cochran}
\affiliation{Department of Astronomy and McDonald Observatory, University of Texas at Austin, 2515 Speedway,~Stop~C1400,~Austin,~TX~78712,~USA}
\author{Luc\'ia Gonz\'alez Cuesta}
\affiliation{Instituto de Astrof\'\i sica de Canarias, C/\,V\'\i a L\'actea s/n, 38205 La Laguna, Spain}
\affiliation{Departamento de Astrof\'isica, Universidad de La Laguna, 38206 La Laguna, Spain}
\author[0000-0002-1887-1192]{Priyanka Chaturvedi}
\affiliation{Th\"uringer Landessternwarte Sternwarte 5, D-07778 Tautenburg, Germany}
\author[0000-0002-0810-3747]{Ilaria Carleo}
\affiliation{Astronomy Department and Van Vleck Observatory, Wesleyan University, Middletown, CT 06459, USA}
\author[0000-0001-6653-5487]{Juan Cabrera}
\affiliation{Institute of Planetary Research, German Aerospace Center, Rutherfordstrasse 2, 12489 Berlin, Germany}
\author[0000-0003-4745-2242]{Paul G. Beck}
\affiliation{Institut f\"ur Physik, Karl-Franzens Universit\"at Graz, Universit\"atsplatz 5/II, NAWI,  Austria}
\author[0000-0003-1762-8235]{Simon Albrecht}
\affiliation{Stellar Astrophysics Centre, Dep. of Physics and Astronomy, Aarhus University, Ny Munkegade 125, DK-8000 Aarhus C, Denmark}
\begin{abstract}
$\pi$~Men hosts a transiting planet detected by the   TESS space mission and an outer planet in a 5.7-yr orbit discovered by  RV surveys. We studied this system using new radial velocity (RV) measurements  taken with the HARPS spectrograph  on  ESO's 3.6-m telescope as well as archival data. We   constrain the stellar RV semi-amplitude due to the transiting planet, $\pi$~Men\,c, as $K_\mathrm{c}$\,=\,1.21\,$\pm$\,0.12\,\ms resulting in a  planet mass of $M_\mathrm{c}$\,=\,3.63\,$\pm$\,0.38\,M$_\oplus$. A  planet radius of $R_\mathrm{c}$\,=\,2.145\,$\pm$\,0.015\,R$_\oplus$ yields a bulk density of $\rho_\mathrm{c}$\,=\,2.03\,$\pm$\,0.22\,g\,cm$^{-3}$. The precisely determined density of this planet and the brightness of the host star make $\pi$~Men\,c an excellent laboratory for internal structure and atmospheric characterization studies. Our HARPS RV measurements also reveal compelling evidence for a third body, $\pi$~Men\,d, with a minimum mass $M_\mathrm{d}$\,sin\,$i_\mathrm{d}$\,=\,13.38\,$\pm$\,1.35\,M$_\oplus$ orbiting with a period of $P_\mathrm{orb,d}$\,=\,125~d on an eccentric orbit ($e_\mathrm{d}$\, = 0.22). A simple dynamical analysis indicates that the orbit of $\pi$~Men\,d is stable on timescales of at least 20 Myrs. Given the mutual inclination between the outer gaseous giant and the inner rocky planet and the  presence of a third body at 125\,d, $\pi$\,Men is an important planetary system for dynamical and formation studies. 
\end{abstract}

%% Keywords should appear after the \end{abstract} command. 
%% See the online documentation for the full list of available subject
%% keywords and the rules for their use.

%% From the front matter, we move on to the body of the paper.
%%
%% We recommend that authors also use the natbib \citep
%% and \citet commands to identify citations.  The citations are
%% tied to the reference list via symbolic KEYs. The KEY corresponds
%% to the KEY in the \bibitem in the reference list below. 

\section{Introduction}
\label{sec:intro}

The bright ($V$\,=\,5.65; Table~\ref{tab:star_parameters}) G0 dwarf star $\pi$\,Men has been the target of exoplanet studies  for over 20 years. \citet{2002MNRAS.333..871J} reported long period ($P_\mathrm{orb,b}$ $\approx$ 2100 d) radial velocity (RV) variations that were consistent with the presence of a sub-stellar companion ($\pi$\,Men\,b) with a minimum mass of $\approx$\,10\,$M_\mathrm{Jup}$ in a highly eccentric ($e_\mathrm{b}$\,$\approx$\,0.6) orbit. The star was later found by NASA's Transiting Exoplanet Survey Satellite \citep[TESS,][]{2014SPIE.9143E..20R} to host a small planet ($R_\mathrm{c}$\,$\approx$\,2\,R$_\oplus$) in a 6.27-d orbit \citep[$\pi$\,Men\,c;][]{2018ApJ...868L..39H,2018A&A...619L..10G}. Due to the brightness of the host star, this planet is a prime candidate for atmospheric characterization studies \citep{2020ApJ...888L..21G,2021ApJ...907L..36G}.

Accurate planetary masses and radii are important for exoplanet studies. True planet masses are needed for understanding the architecture of exoplanets and for dynamical studies. Accurate bulk densities are essential for constraining the internal composition of the planet. In the case of $\pi$~Men~c,  the measured RV amplitude (and thus mass) was based on archival data taken with the HARPS spectrograph. The observing strategy  consisted of  sparse observations spread over a relatively long time and thus was not geared to detect small, short period planets. The corresponding error in the mass, $\approx$\,20\,\%, made it difficult to distinguish between interior models consisting mostly of water, or having a significant fraction ($\approx$\,50\,\%) of silicates \citep{2018A&A...619L..10G}.

The $\pi$ Men system has received heightened interest because  astrometric studies have determined the orbital inclination of the outer planet. \citet{2020MNRAS.497.2096X}, \citet{2020A&A...640A..73D}, and \citet{Damasso2020} combined the long time series of RV measurements for this star along with Hipparcos and Gaia astrometric measurements to determine an orbital inclination of $i_b$ $\approx$ 50$^{\circ}$. Not only  does this pin down the true planet mass as $M_\mathrm{b}$\,$\approx$\,14~M$_\mathrm{Jup}$, but more importantly it establishes that the inner and outer planetary orbits are misaligned. $\pi$~Men joins $\upsilon$ Andromedae  \citep{2010ApJ...715.1203M} and Kepler-108  \citep{2017AJ....153...45M} as stars hosting planets with large mutually inclined orbits. The $\pi$~Men system is thus important for constraining planet formation theories and studying the dynamical evolution of planetary systems.  

In order to measure a more precise mass for $\pi$\,Men\,c,  we included observations of the host star as part of our ESO HARPS large program of spectroscopic follow-up of transiting planet candidates found by TESS. The purpose of these RV measurements was twofold. First, we wished to improve on the error in the planet mass to constrain better compositional models. An accurate mass is also needed to estimate the atmospheric pressure scale height needed for planning observations to detect atmospheric features. Second, we wished to study the architecture of this system by searching for additional planetary companions. This is of especial importance given the recent discovery of the mutual inclination of the  orbits between the inner and outer planets.

\begin{figure}[ht!]
%\plotone{PiMen_b.pdf}
\includegraphics[trim = -4cm 0cm 0cm 2cm clip, scale=.55]{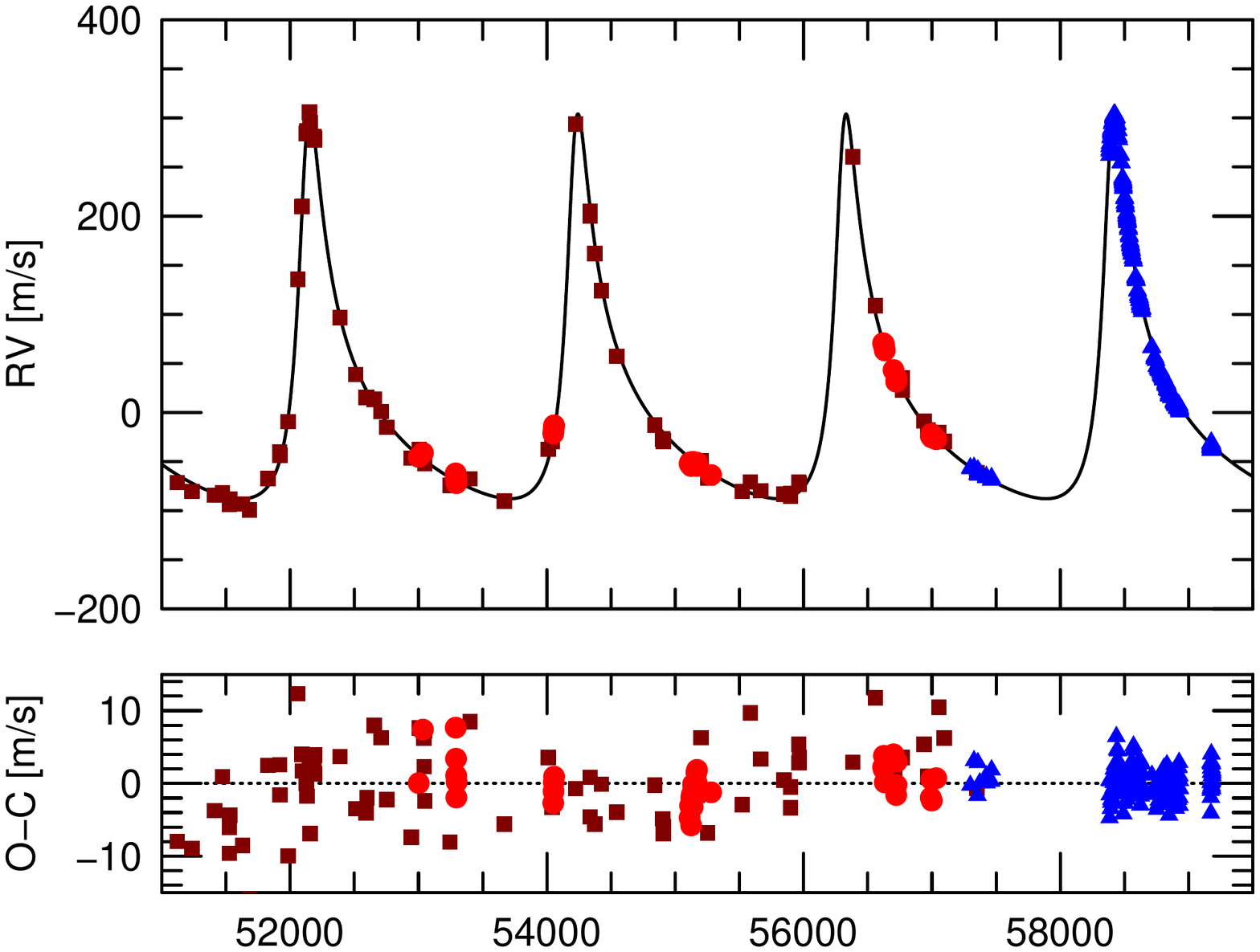}
\caption{(Top) The RV time series from UCLES (brown squares), HARPS-PRE (red dots) and HARPS-POST (blue triangles). For clarity we do not plot the ESPRESSO measurements  as these are contemporaneous with the HARPS-POST data. The curve is the Keplerian orbital solution for $\pi$\,Men\,b. (Bottom) The residuals after subtracting the orbit of $\pi$\,Men\,b.
 \label{fig:PiMen_b}}
\end{figure}

$\pi$~Men was also intensively
observed by the newly commissioned ESPRESSO spectrograph on the ESO's VLT which yielded a $K$-amplitude of 1.5\,$\pm$\,0.2~\ms\ for the inner transiting planet $\pi$\,Men\,c \citep{Damasso2020}. This result offered us a chance to compare the performance of two state-of-the-art spectrographs designed for precise RV work, one a venerable instrument, HARPS \citep{Mayor2003}, mounted on a 3.6-m telescope and in use for almost 20 years, and a more modern one, ESPRESSO, mounted on an 8.2-m telescope \citep{Pepe2014}. The star is bright, so high signal-to-noise ratio ($S/N$) data can be obtained on both instruments with relatively short exposure times. In the future, considerable telescope resources will be invested in the RV follow-up of transiting small planets found by the PLATO mission \citep{2014ExA....38..249R}, so it is useful to compare the performance of both instruments using contemporaneous observations on the same target.

In this work, we adopted the most recent stellar parameters derived by \citet{Damasso2020}. The only exception is the stellar radius, which comes from Csizmadia et al. (2021, submitted). For the sake of completeness, they are listed in Table~\ref{tab:star_parameters}, along with the equatorial coordinates, V-band magnitude, parallax, distance, and proper motion of the star.

\section{The Radial Velocity Data}

Our RV data consists of archival as well as new measurements from our ESO's HARPS large follow-up program. A total of 77 RV measurements come from the UCLES spectrograph mounted on the 3.9-m telescope of the Anglo Australian Telescope (AAT). These can be found in   \citet{2006ApJ...646..505B}  or in \citet{2018A&A...619L..10G}. R. Wittenmyer kindly provided us with the more recent measurements used by \citet{2018ApJ...868L..39H}. Archival HARPS data consisting of 145 RV measurements (51 nightly averaged) were also taken from the  public archive of the European Southern Observatory (ESO). We also included 275 RV measurements (37 nightly averaged) taken with the ESPRESSO spectrograph  \citep{Damasso2020}.

The new data for $\pi$ Men were taken as part of our ESO observing programs 0101.C-0829, 1102.C-0923, and 106.21TJ.001 (PI: Gandolfi) and during technical nights (program IDs 60.A-9700 and 60.A-9709) on HARPS. These consisted of 413 RV measurements\footnote{We excluded 7 measurements taken on 27 and 28 November 2018 (UT), which are affected by a poor wavelength solution \citep[see][]{Nielsen2020}. We also rejected 2 spectra taken on 5 December 2019 (UT), owing to a failure of the telescope guiding system. This gives 404 useful HARPS spectra.}  spanning September 2018 to December 2020 (hereafter ``HARPS-Large'' data set). The star is bright, so we typically took several observations per night. Exposure times were 150-300 secs resulting in a median signal-to-noise ratio (S/N) of $\sim$250 per pixel at 550\,nm. If one considers only nightly averages our program resulted in 177 new measurements taken at different epochs. In June 2015 the HARPS fiber bundle was upgraded \citep{2015Msngr.162....9L}. This results in a zero-point offset between the data taken before and after the upgrade. We therefore treated the complete data as  four independent sets with different zero point offsets: UCLES, ESPRESSO and HARPS before (HARPS-PRE) and after the fiber upgrade (HARPS-POST, this set also includes HARPS-Large).

We extracted the HARPS spectra using the Data Reduction Software \citep[DRS;][]{Lovis2007}. There are a number of reduction pipelines available for the calculation of RVs: The DRS, which uses the cross-correlation method with a digital mask \citep{Pepe2002}, the HARPS Template-Enhanced Radial velocity Re-analysis Application \citep[HARPS-TERRA,][]{2012ApJS..200...15A} and the SpEctrum Radial Velocity AnaLyser (SERVAL) pipeline \citep{2018A&A...609A..12Z}. For our final analysis we used the RVs calculated with HARPS-TERRA as this produced a final root mean square (rms) scatter that was about 4\,\% lower than the other two methods. We emphasize, however, that {\it all} reduction programs produced consistent orbital parameters that were well within the uncertainties.

Briefly, HARPS-TERRA performs a least squares matching of each observed spectrum to a high signal-to-noise template spectrum produced by co-adding all the observations of the target star after they have all been placed on the same wavelength scale and corrected for the Earth's barycentric motion. The RV for each spectrum is calculated with respect to this master template.

$\pi$ Men is a high proper motion star. The RV data span 17 years so it is important to remove the secular acceleration. The star has a systemic radial velocity of 10.595\,$\pm$ 0.0003\,km\,s$^{-1}$ (as derived from the analysis of the HARPS DRS RVs), a parallax of 54.705\,$\pm$\,0.067 mas, and proper motion of 311.187\,$\pm$\,0.127\,mas\,yr$^{-1}$ and 1048.845\,$\pm$\,0.136\,mas\,yr$^{-1}$ in RA and Declination, respectively \citep[][see also Table~\ref{tab:star_parameters}]{GaiaDR2}. This results in a secular acceleration of $\sim$0.48 m\,s$^{-1}$yr$^{-1}$. This was removed for all RVs except for those processed with HARPS-TERRA, which already accounts for the secular acceleration in its pipeline.

\begin{figure}[ht!]
\plotone{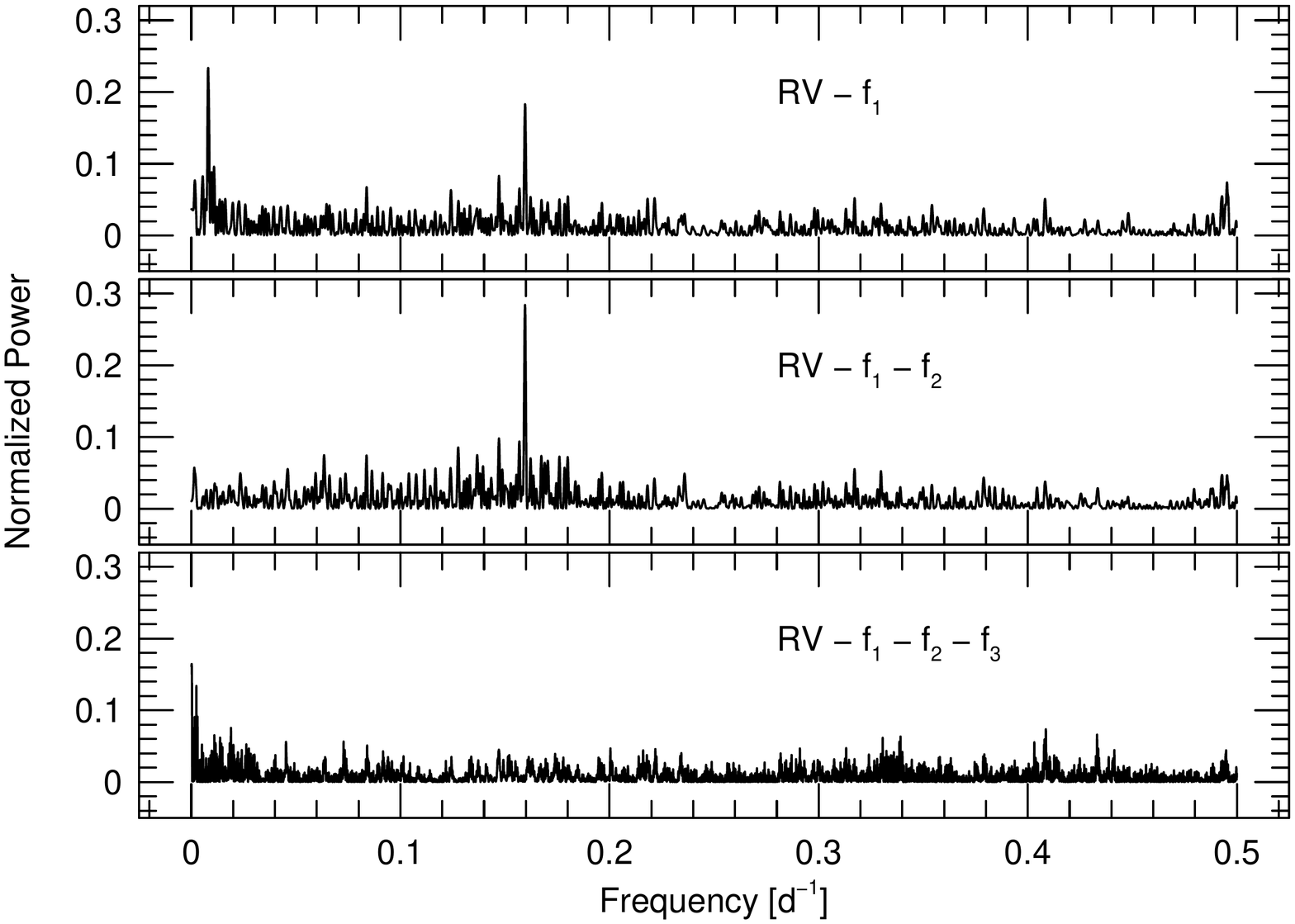}
\caption{(Top) The GLS periodogram of the HARPS-Large RV measurements after removing the motion of the outer planet
(orbital frequency, $f_1$ = 4.79 $\times$ 10$^{-4}$ d$^{-1}$). (Middle) The GLS periodogram of the RV residuals after removing the contribution of the dominant peak at $f_2$ =
0.008 d$^{-1}$. (Bottom) The GLS periodogram of the final RV residuals after removing the orbital frequency of the transiting planet, $f_3$ =\,0.16\,d$^{-1}$.
\label{fig:GLS}}
\end{figure}
    
Table~\ref{tab:rv_sets} lists the data sets used in our RV analysis. We reiterate that the ``HARPS-POST'' data contains both the 17 archival HARPS measurements after the upgrade and those from our large program. The ``HARPS-Large'' is a subset of this which only contains our new 404 useful measurements from the large program. The standard deviation listed is the root mean square (rms) scatter of the data sets after the removal of all periodic signals (see below). Table~\ref{tab:rv_data} lists the new RV measurements from our ESO large programs.

\section{Periodogram Analysis of the RV Data}

We first  performed a frequency (periodogram) analysis on the RV data in order to confirm  that the signal of the transiting planet $\pi$ Men~c is indeed present in the data. Furthermore, identifying all significant signals in the data and modeling these is important for deriving a precise RV amplitude for the transiting planet. To find weak periodic signals we first had to remove the large variations due to the outer planet. Since the UCLES measurements increase the time base of the measurements needed to refine the parameters of the outer planet, we included these in spite of these having  poorer RV precision.  

An initial orbital solution
%\footnote{Later, we will perform a a joint fit using all known periodic signals in the RV time series.} 
was performed using the general non-linear least squares fitting program
{\it Gaussfit} \citep{1988CeMec..41...39J}. All orbital parameters were allowed to vary, including the zero-point offset of each individual data set. The orbit parameters listed in Table~\ref{tab:orbits} represent the final ones from our joint fit (see below), which are entirely consistent with the {\it Gaussfit} results. The orbital fit for $\pi$\,Men\,b is shown in Figure~\ref{fig:PiMen_b}. The UCLES measurements have an rms scatter more than twice that of the HARPS data and will be excluded from the subsequent analyses. 

%\begin{figure}[ht!]
%\plotone{p121hist.pdf}
%\caption{The distribution of the peak in the periodogram of simulated data using an input period of 121 d.}
 %\label{fig:p121histogram}
%\end{figure}

\begin{figure}[ht!]
\plotone{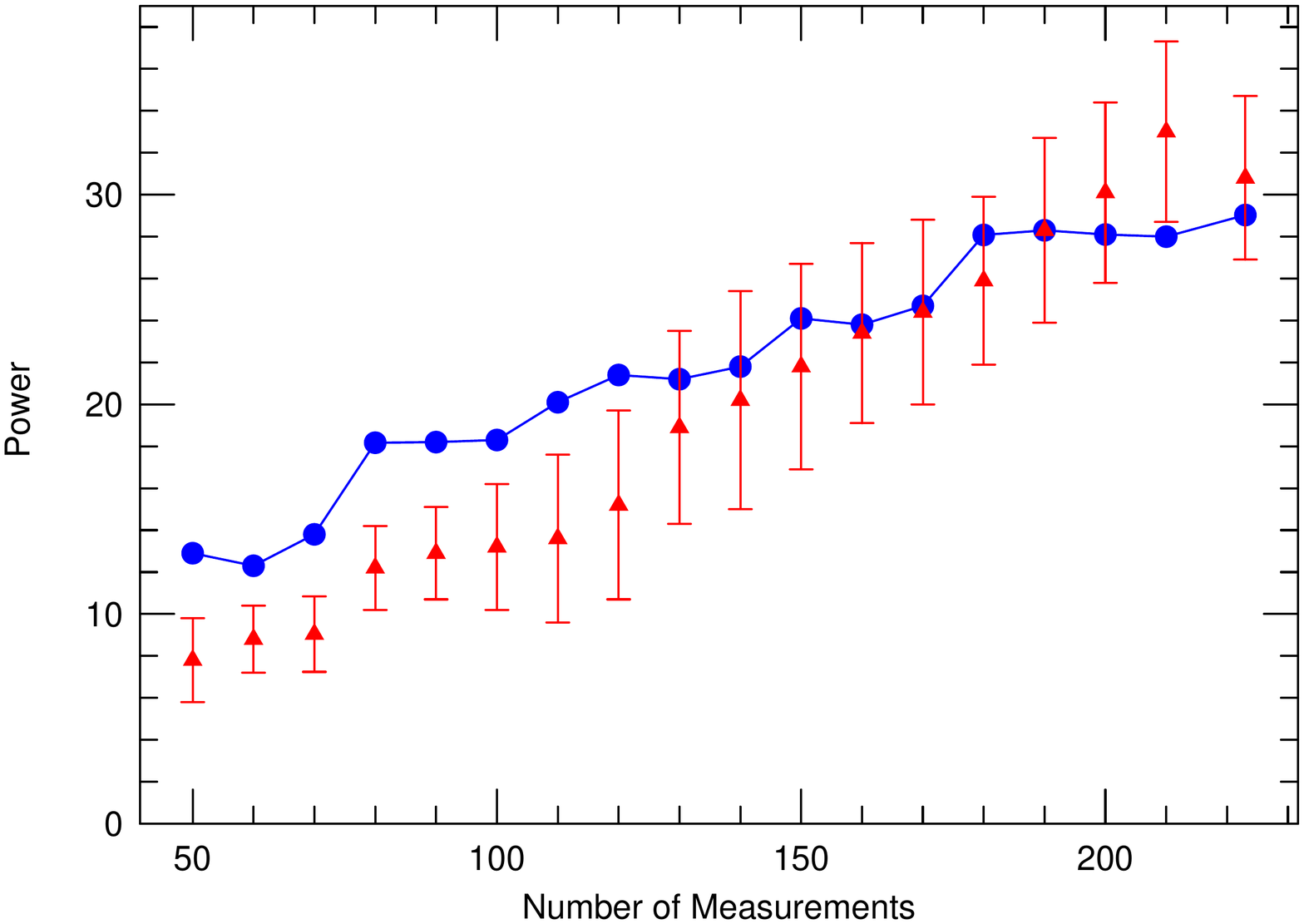}
\caption{Growth of the Lomb-Scargle power of the 125-d period attributed to a third body in the system as  a function of the number of RV measurements for the real data (HARPS-Large and ESPRESSO, blue dots) and simulations (red triangles). 
 \label{fig:planet_d_pow}}
\end{figure}

\begin{figure}[ht!]
\plotone{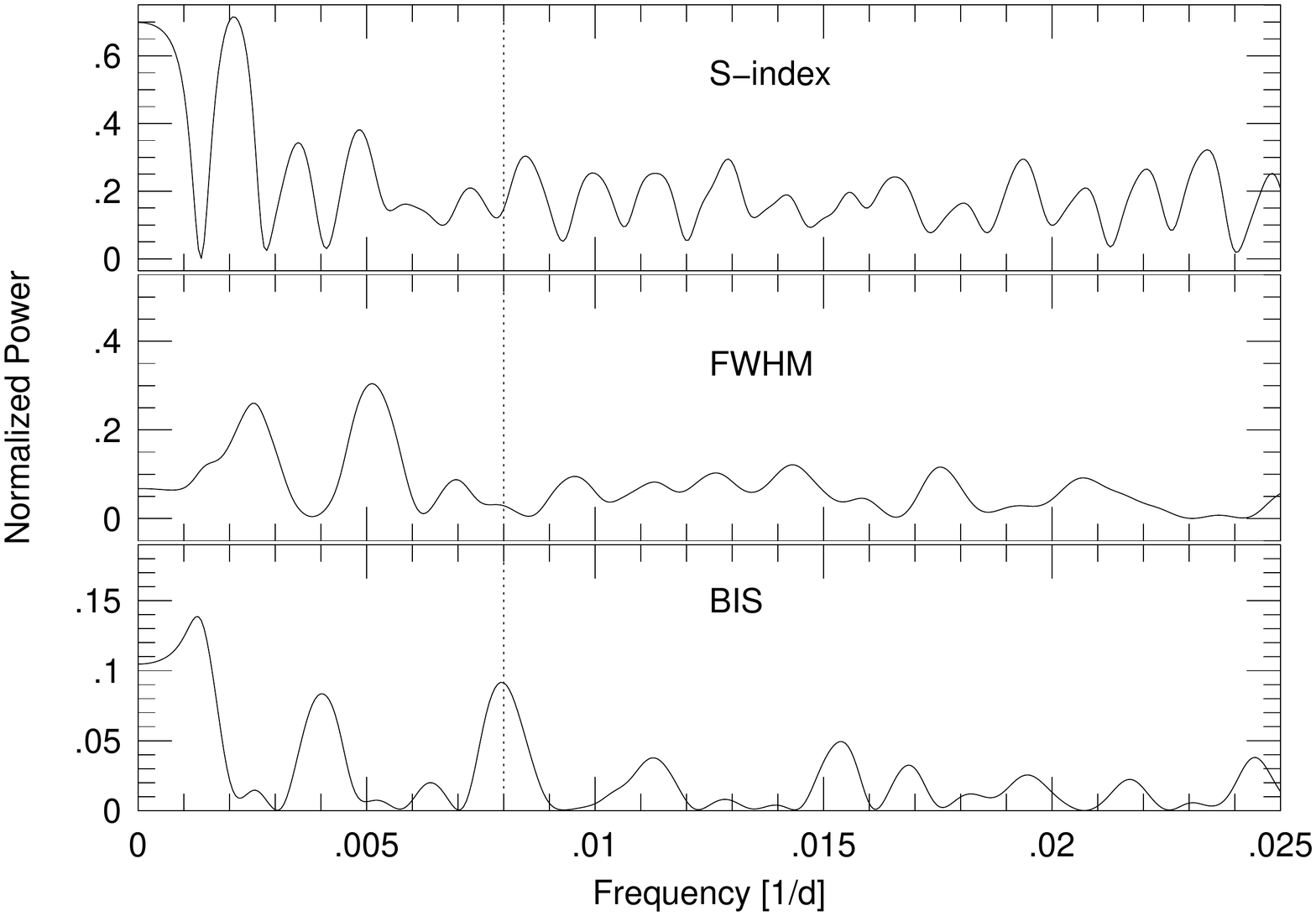}
\caption{The GLS periodograms of the activity indicators extracted from the HARPS-Large data-set. From top to bottom: S-index, FWHM, and  BIS. The vertical dashed line marks the
orbital frequency of the 125-d  detected in the RVs.}
 \label{fig:activity_gls}
\end{figure}

\begin{figure}[ht!]
\plotone{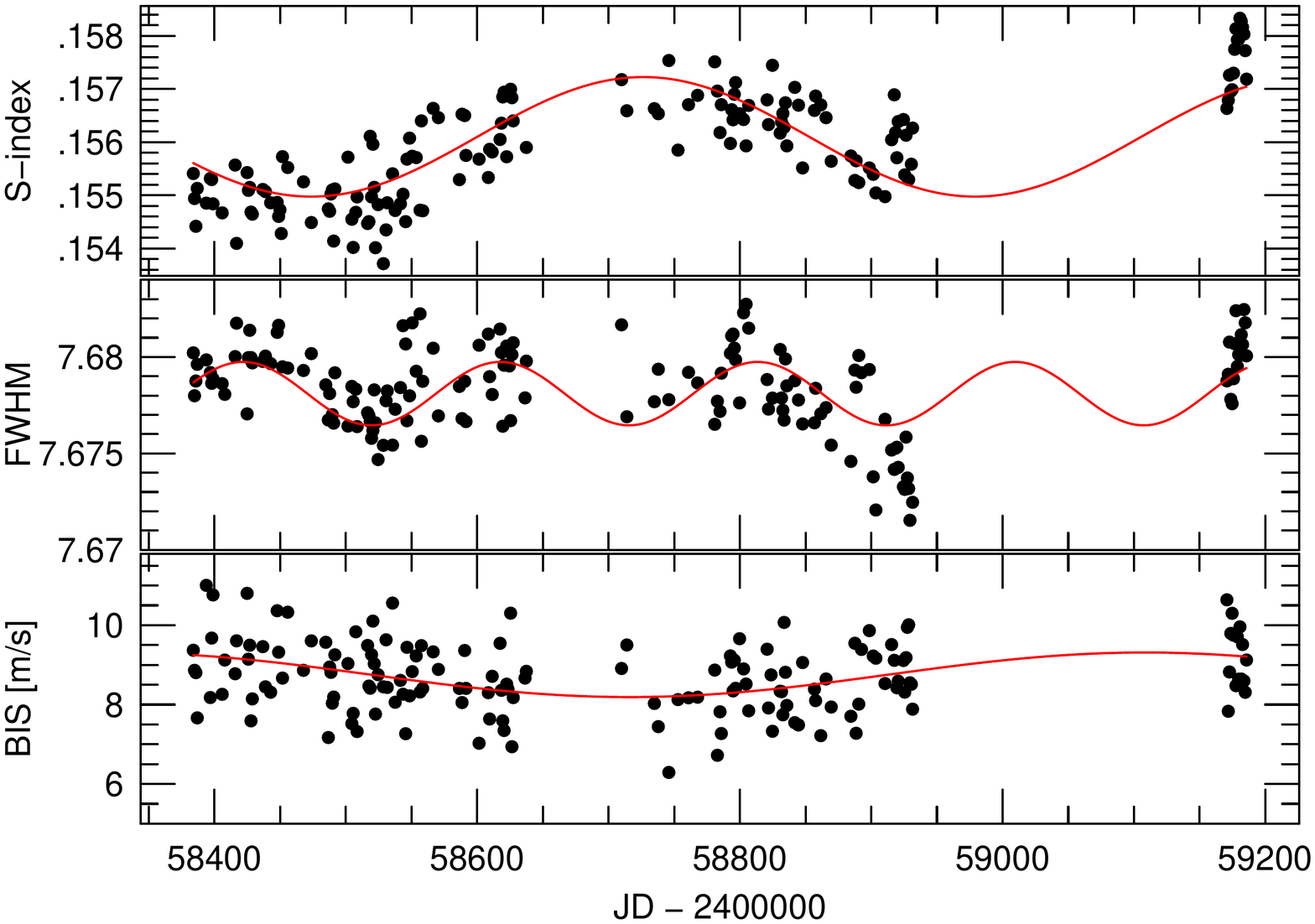}
\caption{The time series of the S-index (top), FWHM (middle) and BIS (bottom) measurements extracted from the HARPS-Large data-set. The red curves are sine fits using the period
 of the dominant peak in each periodogram (Table 5).}
\label{fig:indicators_time}
\end{figure}

\begin{figure}[ht!]
\plotone{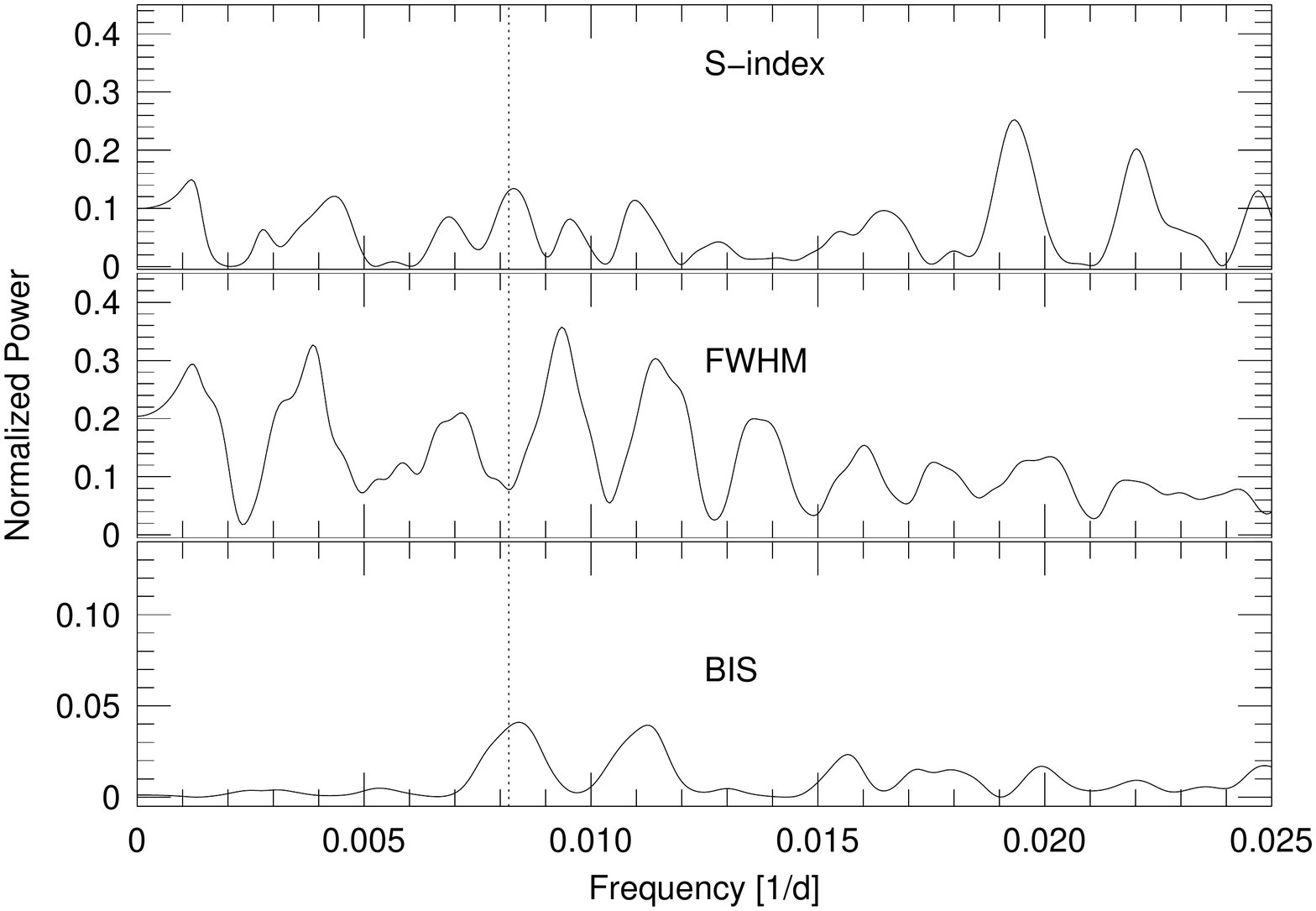}
\caption{GLS periodogram of the activity indicators after removal of the dominant signal in each (Table~\ref{tab:indicators}).
The veritical  
dashed line is the frequency of the  125-d period seen in the RV data. }
 \label{fig:activity-f1}
\end{figure}

The generalized Lomb-Scargle (GLS) periodogram  \citep{2009A&A...496..577Z} of the residual HARPS-Large RV data, after removing the orbital frequency  of planet b
($f_1$ = 4.79 $\times$ 10$^{-4}$ d$^{-1}$), is shown in the top panel of Figure~\ref{fig:GLS}. We only used the HARPS-Large data for this analysis for two reasons. First, it provides a much simpler sampling window. Including the early HARPS-PRE data results in a very complex window with many more alias peaks. Second, the HARPS-PRE data start approximately 10 years earlier and have much sparser sampling. This means that any underlying long term stellar variability, which will be difficult to model, can boost power into an alias frequency thus masking the true one that is present. 

The  highest peak occurs at a frequency of $f_2$\,=\,0.008 d$^{-1}$ ($P$\,=\,125\,d), although one can see significant power at the orbital frequency of the transiting planet ($f_3$\,=\,0.16\,d$^{-1}$; P\,=\,6.27 d). Removing $f_2$ increases the power seen at orbital frequency of the transiting planet (middle panel of Figure~\ref{fig:GLS}). The final residuals (lower panel) results in no additional significant peaks. A peak is seen at $f$\,=\,1.56\,$\times$\,10$^{-3}$ ($P$\,=\,640 d), but this has low significance with an estimated false alarm probability, FAP\,$\approx$\,2\,\%, which we do not consider 
significant.

We assessed the statistical significance of the new $\sim$125\,d period via the bootstrap randomization process where the RV values were randomly shuffled keeping the time stamps fixed \citep{1993ApJ...413..349M}. In 300,000 realizations of the {bootstrap} there was no instance where the random data periodogram showed power higher than the real data. This implies a FAP  $\ll$\,3.3\,$\times$\,10$^{-6}$. It is highly unlikely that this peak is due to random noise. 

%\begin{figure}[ht!]
%\plotone{RHKphase.pdf}
%\caption{The residual R'HK (top) and RV (bottom) meausurements
%phased to the 125-d period. The red line is a sine fit of the data.
%}
%\label{fig:RHKphase}
%\end{figure}

\begin{figure}[ht!]
\plotone{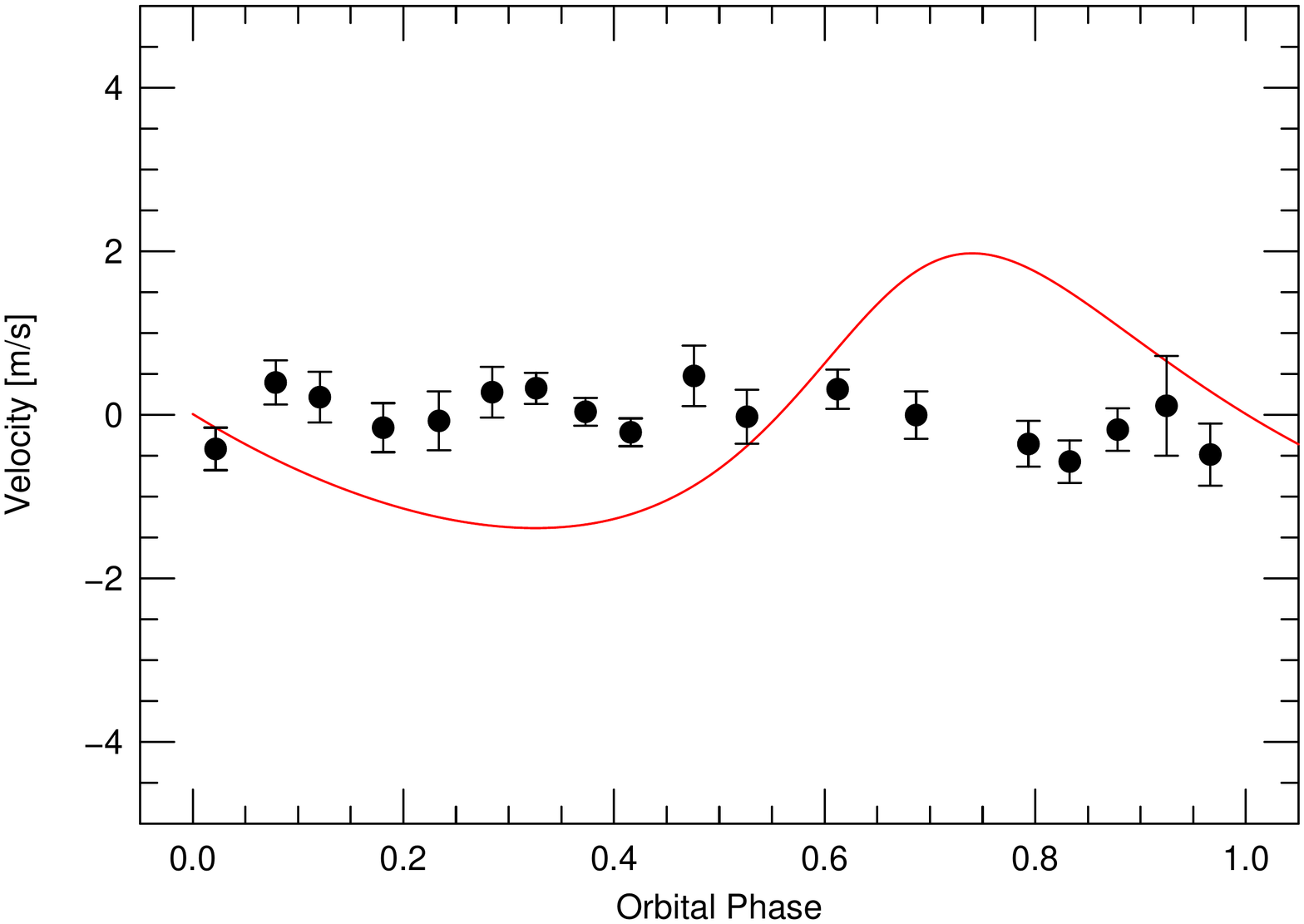}
\caption{The phase-binned BIS measurements using the $\approx$ 125-d period found in the RVs. The red curve represents
an orbital solution to the 125-d period (see Section 5.1).}
\label{fig:bis_phase}
\end{figure}

\section{The Nature of the 125-d Period}

 Before we can adequately model the 125-d signal in our analysis, it is important  to establish its nature. If it is planetary in origin it would be an important new member of the $\pi$~Men system;  however, it can also arise from stellar activity.  Important criteria for establishing the planetary nature of an RV signal are that it is long-lived and coherent and that it is not present in any activity indicators.

\subsection{Coherence of the 125-d Period}

There are several ways to test whether the 125-d period in the RV data  is coherent and stable.  One can examine the evolution of the power in the standard Lomb-Scargle (LS) periodogram as a function of the number of measurements \citep{2016A&A...585A.144H} or the evolution of the false alarm probability \citep{2018A&A...609A.117T}. Alternatively, one can use the Stacked-Bayesian GLS periodogram \citep{2017A&A...601A.110M} to assess the stability of a signal. However, it can be difficult to assess the stability of  a signal merely by looking at the stacked periodogram. The pathology of the sampling often can make a signal appear unstable for a time when in fact it is not,  or vice versa. We therefore chose to  use the growth of the LS power and compare it to what is expected from  a simulated signal with the appropriate noise added. This provides a somewhat more ``quantitative'' assessment of the stability. 

We first isolated the 125-d signal by removing the contribution of the outer and transiting planets. The blue dots in Figure~\ref{fig:planet_d_pow} shows the growth of the LS power ($P$) of the 125-d signal as a function of the number ($N$) of RV measurements (we will refer to this  as the $P$-$N$ relationship) using the HARPS-Large and ESPRESSO data. We then compared this to expectations (red triangles) using an orbital fit to the 125-d RV variations (see below) sampled like the data and with the appropriate random noise added ($\sigma$ = 1.5\,m\,s$^{-1}$). The error bars indicate the standard deviation in the power using simulated data with 10 different realizations of the noise. The evolution of the power of the 125-d period largely follows the simulated data in that there is a monotonic increase in the power as a function of number of data points, although the slope in the power evolution of the real data is slightly shallower than the simulated data. Overall, the behavior of the $P$-$N$ 
relationship seems to indicate that the 125-d period is stable and coherent.

\begin{figure}[ht!]
\plotone{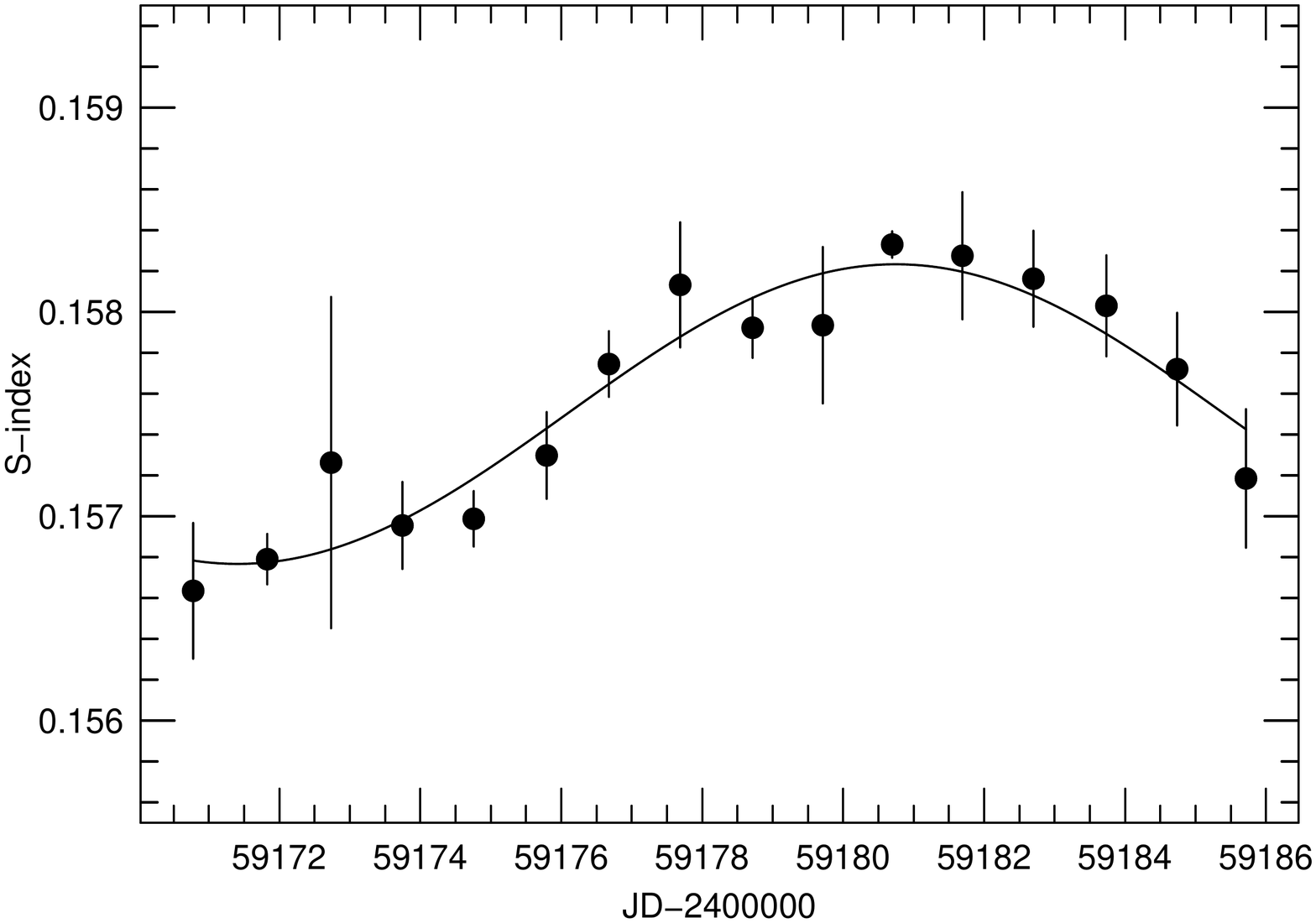}
\caption{S-index measurements over 16 consecutive nights. The curve represents the sine fit with a period of 18.7 d.
}
\label{fig:rotation}
\end{figure}

\subsection{Activity Indicators}

Stellar activity can also create periodic RV signals that can mimic a planet and this can be revealed by a periodogram analysis of activity indicators.  If an RV signal is absent in all  activity indicators then we can be more confident that it is the barycentric motion of the host star due to the presence of a companion. We have three activity indicators at our disposal: The full width at half maximum (FWHM) and bisector inverse slope (BIS) of the cross-correlation function (CCF), as well as the Ca\,{\sc II} S-index measurement \citep[see][]{1995ApJ...438..269B}. HARPS-TERRA also delivers  indices on the Balmer H$\alpha$ and Na D lines. However, a period analysis of these data showed  a dominant period at one year, possibly an indication of telluric contamination. We therefore excluded these from our analysis.

Figure~\ref{fig:activity_gls} shows the GLS periodograms for the three activity indicators extracted from HARPS-Large data. All have the highest peak at low frequencies that seem to be unrelated to the frequency found in the RV time series. The only exception is the BIS which shows a secondary peak near 0.008\,d$^{-1}$ (see Sect.~\ref{Sec:Bis_Var} for a more detailed discussion of the bisector variations). Table~\ref{tab:indicators} lists the dominant periodic signals found in the activity indicators and the FAP determined via 200,000 realizations of a bootstrap. Figure~\ref{fig:indicators_time} shows the fit to the time series of each activity indicator using the period of the dominant peak found in the corresponding periodogram. 

We investigated whether additional signals could be hidden in the activity indicators. Pre-whitening, i.e., the subsequent fitting and removal of dominant frequencies in the periodogram of the time series and its residuals, is a powerful tool for finding weak, underlying signals in data. For instance, pre-whitening of the H$\alpha$ index measurements for GL\,581 was able to isolate variations with the same period as the purported planet GL\,581\,d  \citep{2016A&A...585A.144H}, thus supporting the claim of activity as the origin for the 66-d RV period \citep{2014Sci...345..440R}.

The GLS periodograms of the residuals of the activity indicators, after removing the appropriate sine fits to the dominant frequency in the data, are shown in Fig.~\ref{fig:activity-f1}. The S-index and FWHM residuals do not show any significant\footnote{We considered a signal to be significant if its false alarm probability is FAP\,$<$\,0.1\,\%.} peaks at 0.008\,d$^{-1}$. The S-index only shows a weak one that is the fifth highest peak in the frequency range.  In the amplitude spectrum it appears less than twice the mean height of the surrounding noise peaks and indication that it is not significant  \citep[see][]{1997A&A...328..544K}.

The peak in the BIS residuals at 0.008\,d$^{-1}$ is the highest peak in the frequency range considered (Fig.~\ref{fig:activity-f1}).  We assessed the significance of this peak via a bootstrap. The FAP that noise can produce a peak {\it exactly} coincident at this frequency is about 7\,\%, which we do not consider statistically significant. Note that a nearby peak has comparable power.

 \subsubsection{The Bisector Variations}
 \label{Sec:Bis_Var}
 
The bisector is the only activity indicator that shows possible variations at the 125-d RV period.  Both the raw and residual BIS variations show an isolated and modestly strong peak at the period (frequency) of interest. Since the bisector variations may be  the only indicator to cast doubt on a planet hypothesis for the RV period and given the importance of establishing that the 125-d period is planetary in nature,
this merits a critical evaluation of the reality of these.

The periodogram of the raw bisector measurements (Fig.~\ref{fig:activity_gls}, lower panel) shows a peak coincident with the 125-d period (f\,=\,0.008 d$^{-1}$). Without any a priori information this has a false alarm probability of  FAP\,$\sim$\,30\,\%. However, this FAP represents the probability that noise can create a peak at least this high over a broad frequency range. As pointed out by \citet{1982ApJ...263..835S}, for a known signal (i.e., 125\,d) we need to consider the probability that noise can create a peak with the observed power or higher {\it exactly} at this frequency. If $z$ represents the un-normalized power, then FAP $\sim$ $e^{-z}$. In this case the FAP is rather low at  FAP\,=\,0.13\%. This value is consistent the FAP obtained with a bootstrap analysis. In spite of this low FAP we are confident that this signal is not significant for several reasons.

First, the bisector data are quite noisy. A sine fit to these using the 125-d period results in an amplitude of 0.33~\ms\ or 2.5 times smaller than the rms scatter ($\sigma$\,=\,0.83~\ms) about the fit. In spite of this, the 125-d signal should have been detected at much higher significance due to the large number of measurements. As a test, we took a 125-d sine function with an amplitude of 0.33~\ms\ and added random noise with the rms scatter of the measurements. The periodogram of the simulated data 
showed power consistent with a FAP that was more 
than 1000 times lower than was seen for the real data. A true signal should have shown a much higher power (i.e. lower FAP) than is observed.

Second, the pre-whitened analysis provides unconvincing support that the bisector variations are intrinsic to the star. Our analysis of the residuals showed a false alarm probability of FAP\,$\sim$7\,\%.  Since we were focused on the 125-d signal we initially only considered the frequency range out to 0.025\,d$^{-1}$  (Fig.~\ref{fig:activity-f1}, lower panel). If we extend the analysis to higher frequencies we find that the highest peak in the residuals occurs at 20\,d (f\,=\,0.05\,d$^{-1}$). Removing this signal weakens the peak at 125-d further such that its FAP increases to $\sim$50\,\%.

{Third, we examined whether any  phase variations of the bisectors can be related to the 125-d signal. Figure~\ref{fig:bis_phase} shows the orbital fit to the 125-d RV variations. This fit is presented and discussed in detail below in Section 5.1. The points show the binned (bin size $\approx$ 0.05 in phase) residual  BIS
measurements, after removing the dominant signal,  phased to the same period. There is no hint of sinusoidal variations in the BIS that could account for the observed 125-d RV signal. This strongly indicates that the signal found in the periodogram is due to noise, consistent with the large FAP we derived.}

 Finally, it is worth noting the case of 51\,Peg which highlights the pitfalls of relying only on bisector measurements to refute a planet hypothesis. \citet{1997ApJ...490..412G} found bisector variations in 51\,Peg that were coincident with the orbital period of the planet. The formal FAP for this signal was 0.3 \% and thus seemed to be significant. Subsequently, higher quality bisector measurements were found to be constant \citep{1998Natur.391..154H}. In spite of the low FAP, the previous variations were in fact due to noise. Since the bisector signal at 125 d in $\pi$ Men is not supported by other activity indicators we conclude that these do not provide convincing enough evidence to refute a planet hypothesis for the RV signal.

\subsection{The Rotational Period}

In the full time series we found no significant peaks that may be indicative of stellar rotation. This is expected given that rotational modulation from activity may not be coherent over the long time span of our data. However, our time series did have long stretches when closely spaced measurements were made spanning 10-20 days. We searched for evidence of rotational modulation in these subsets and found only  one instance, at the end of the S-index time series, which may show variations due to rotational modulation. Figure~\ref{fig:rotation} shows a 16-d time span of S-index measurements showing sine-like variations with a period of 18.7 $\pm$ 0.8 d. 

\citet{Damasso2020} measured a projected rotational velocity of the star of $v$\,sin\,$i$\,=\,3.34\,$\pm$\,0.07~\kms\ which agrees with the value of 3.3\,$\pm$\,0.5~\kms\ found by \citet{2018A&A...619L..10G}. Csizmadia et al. (2021, submitted) measured a stellar radius of as $R_\star$\,=\,1.190\,$\pm$\,0.004~R$_\odot$, which results in a maximum rotation period of $P_\mathrm{rot}$\,=\,18.0\,$\pm$\,0.4~d, consistent with the value from the S-index variations and that inferred by \citet{Damasso2020}. Interestingly, the 20-d period found in the bisector measurements is close to this value. Given the low significance of the bisector signal we are uncertain that it is truly related to stellar rotation.

\section{Keplerian Motion for the 125-d Period}

The most likely explanation for the 125-d period is that it stems from the presence of a third companion that we refer to as $\pi$\,Men\,d. Here we determine the orbital parameters and investigate the stability of the system.

\subsection{Orbital Solution}

A preliminary Keplerian fit to the 125-d period using the residuals after removing the contributions of the inner and outer planets indicates an eccentric orbit ($e$\,$\sim$\,0.2). We performed a joint fit for all three Keplerian signals using the code \texttt{pyaneti} \citep{Barragan2019}, which employs a Bayesian approach combined with Markov chain Monte Carlo sampling to estimate the planetary parameters. We derived the best-fitting orbital solution for $\pi$\,Men\,b from the UCLES, HARPS-PRE, HARPS-POST, and ESPRESSO date-sets, and used only the HARPS-POST and ESPRESSO date-sets to determine the parameters of $\pi$\,Men\,c and d. We imposed uniform uninformative priors for all the fitted parameters, except for the orbital period and time of first transit of $\pi$\,Men\,c, for which we adopted Gaussian priors from \citet{Damasso2020}. A  circular orbit for the inner transiting planet was assumed,  but we fitted the eccentricity for $\pi$\,Men b and d. We accounted for the RV offsets between the different instruments/setups and fitted for jitter terms to account for both instrumental noise not included in the nominal uncertainties and RV variation induced by stellar activity.  A  parameter space with 500 Markov chains was explored to generate a posterior distribution of 250\,000 independent points for each model parameter. The inferred parameters are given in Table~\ref{tab:orbits} and Table~\ref{tab:PiMen_d}. They are defined as the median and 68\,\% region of the credible interval of the posterior distributions for each fitted parameter. 

%We first derived an orbital solution by fitting the full data set (excluding the AAT RV measurements) and using the best fit 125-d period. The orbital parameters for $\pi$\,Men\,d resulting  from a simultaneous fit to all three planetary signals are listed in Table~\ref{tab:PiMen_d}. 

\begin{figure}[ht!]
\plotone{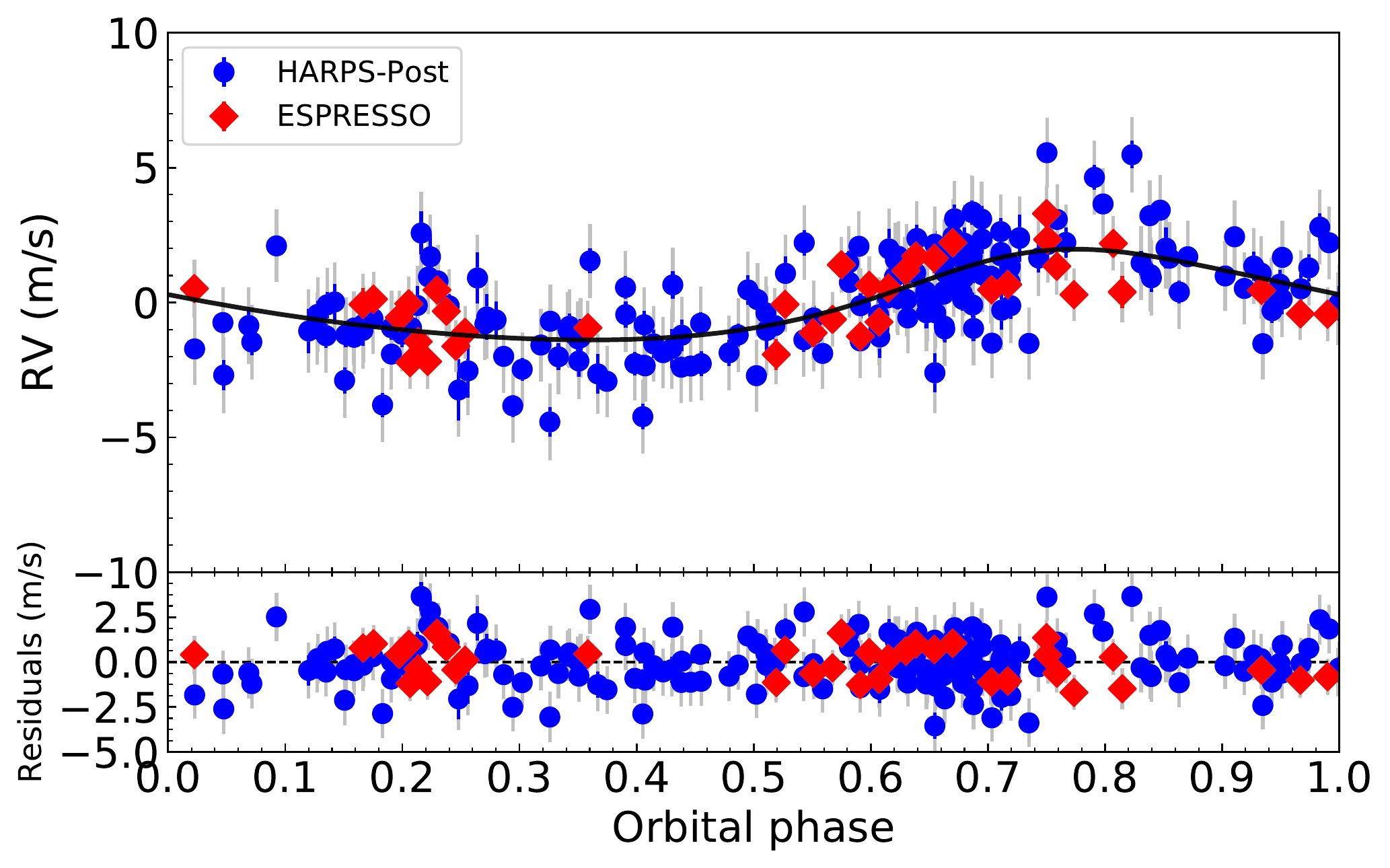}
\caption{RV variations and Keplerian orbit for $\pi$\,Men\,d phased to the orbital period of $\sim$125~d. The contributions from the inner and outer planets have been removed. Blue circles and red diamonds are nightly binned HARPS-POST and ESPRESSO RV measurements, respectively. The  gray  error bars include jitter. The curve represents the orbital solution. The zero phase of the RV curve corresponds to the time of inferior conjunction (see Table\,\ref{tab:PiMen_d}).}
\label{fig:PiMen-d}
\end{figure}

\begin{figure}[ht!]
%\plotone{PiMen-d_time.pdf}
\includegraphics[scale=0.48]{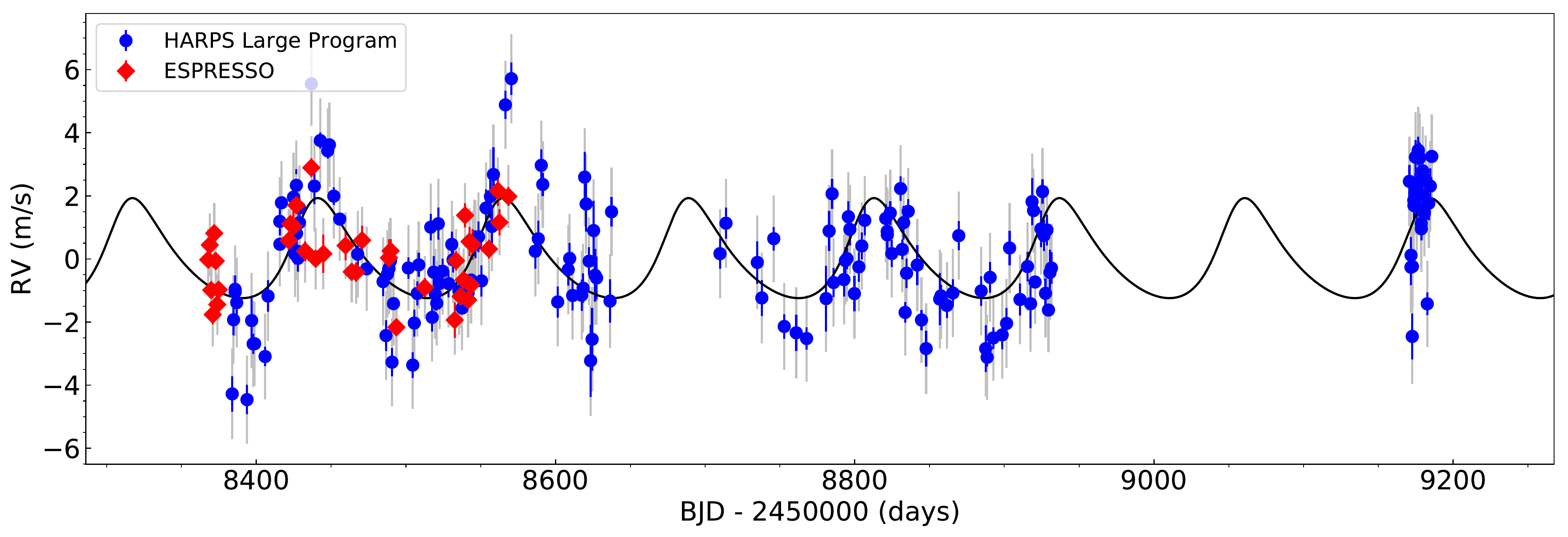}
\caption{The RV versus time from HARPS large program (blue circles) and ESPRESSO (red diamonds) nightly binned measurements for $\pi$\,Men\,d (125-d period) after removing
the contribution of the inner and outer planets. The error   bars include the jitter term. The curve represents the orbital solution.} 
\label{fig:planetd_time}
\end{figure}

$\pi$\,Men\,d has a minimum mass of $M_\mathrm{d}$\,sin$i_\mathrm{d}$\,=\,13.38\,$\pm$\,1.35\,M$_\oplus$ whose orbit is modestly eccentric ($e$\,=\,0.220\,$\pm$\,0.079). Interestingly, the angle of periastron passage, $\omega$ is comparable to that of the outer planet. The phase curve of the orbit is shown in Fig.~\ref{fig:PiMen-d} and the time series, focusing on just the time span of the HARPS-Large RVs, is shown in Fig.~\ref{fig:planetd_time}.

{We found that the difference between the Bayesian information criterion of the 3-planet (b, c, and d) and 2-planet (b and c) models is $\Delta$BIC\,=\,$-$73, providing very strong evidence for the existence of a third Doppler signal in the data}. After removing the contribution of the orbital motions of all planetary signals the HARPS data has an rms scatter of $\sigma_\mathrm{HARPS-Post}$ \,=\,1.40~\ms\ and ESPRESSO, $\sigma_\mathrm{ESPRESSO}$\,=\,0.95\,\ms. Both are about a factor of three larger than  the nominal measurement errors  which are typically better than 0.5~\ms.

\subsection{Orbital Stability}

For the 125-d RV variations to be due to a planet, it must lie on a stable orbit. It is beyond the scope of this paper for a detailed dynamical investigation. Instead, we performed a preliminary analysis to assess if a stable orbit is at least feasible. For this dynamical study we employed  \texttt{Rebound} \citep{2012A&A...537A.128R} and drew samples from the posterior distributions of our 3-planet orbital solutions while the inclination of $\pi$~Men~b was drawn from $i_b$ = 50 $\pm$ 5$^\circ$ according to previous studies 
previous studies (\citealt{2020A&A...640A..73D}; \citealt{2020MNRAS.497.2096X}; \citealt{Damasso2020}). Since we can only derive the minimum mass\footnote{The orbital inclination of the outer planet comes from astrometry and that of the inner planet from the transit light curve.} ($M_\mathrm{d}$\,sin\,$i_\mathrm{d}$) of $\pi$\,Men\,d, we allowed for an orbital inclination between 20$^\circ$ and 90$^\circ$. 

\begin{figure*}[ht!]
%\plotone{dynamic.pdf}
\includegraphics[trim = 7cm 0cm 6cm 0.0cm clip, scale=0.6, angle=90]{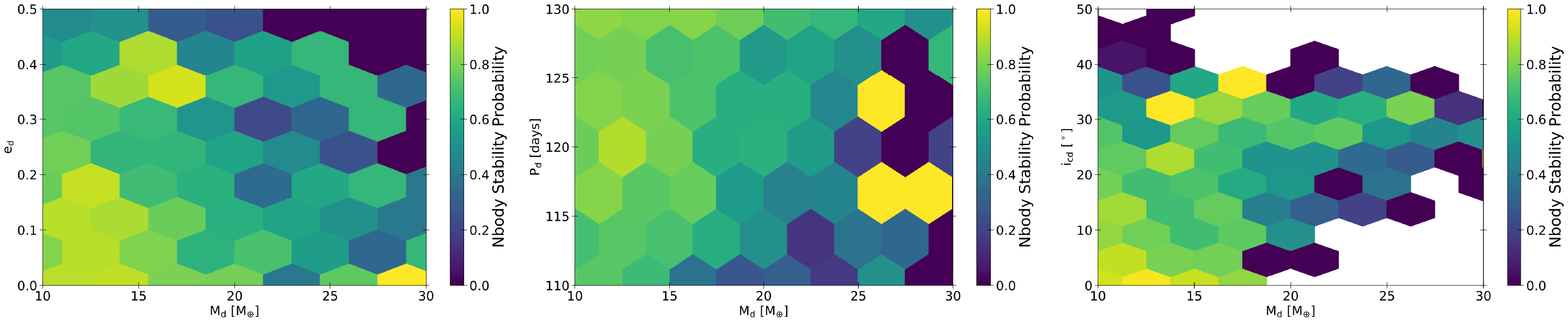}
\caption{Stability maps for $\pi$ Men~d. Yellow and green regions indicate regions with a high probability of stability whereas dark regions are unstable. (Left) Stability regions in the eccentricity -  planet mass plane. (Right) Stability regions in the of orbital period  - planet mass plane.}
\label{fig:dynamic}
\end{figure*}

The left and center panels of Figure~\ref{fig:dynamic} shows the stability probability as a function of the eccentricity, period, and mass of planet d for our simulation. For the  period range of 110-130 d the orbit of planet d is stable, although there are isolated regions where it is unstable. The  planet must have a mass $<$\,20\,M$_\oplus$ for stability, For our orbital solution (Table~\ref{tab:PiMen_d}) this implies an orbital inclination $i_\mathrm{d}$\,$>$\,40$^\circ$. {The right panel of Figure~\ref{fig:dynamic} shows the stability probability
in the mass versus mutual inclination plane. In general orbits are more likely to be unstable for high mutual inclinations and a low mass planet, or high mass even for relatively
low mutual inclinations.}

Due to limited computational resources our simulations only covered a time span of 20\,Myrs,  a small fraction of the estimated stellar age of $\approx$\,3.3 Gyrs \citep{Damasso2020}. Clearly, a numerical simulation is called for covering a much large fraction of the stellar life. For such a simulation covering a longer time span it would be important to obtain more accurate orbital parameters in order to restrict the parameter space of the simulations. 

The left panel of Fig.~\ref{fig:dynamic} shows that the planet's orbit has a higher chance of stability if it lies on a more circular orbit, $e$\,$<$\,0.3 and our measured eccentricity of 0.22 lies below this. Further RV measurements
may reveal that the orbital eccentricity may in fact be lower. For instance,  $\epsilon$ Eri\,b is an exoplanet where the $K$-amplitude is also comparable to the rms scatter of the RV measurements as is the case here. The discovery paper initially reported a high eccentricity ($e$\,=\,0.6) for this planet \citep{2000ApJ...544L.145H}. However, additional measurements spanning 30 years were able to show that the actual orbit was nearly circular, $e$\,= $0.07_{-0.05}^{+0.06}$ \citep{2019AJ....157...33M}. The initial high eccentricity most likely stemmed from the low $K$-amplitude that was comparable to the measurement error and the influence of activity on the RV variations from the orbital motion.

%A more accurate dynamical analysis considering also the secular interactions between the planets and the Kozai effect could be carried out to put tighter constrain on the evolution of the system, but this goes beyond the scope of this work. Furthermore, additional observations are required to further constrain the parameters for the outer planet.

%\begin{figure}[ht!]
%\plotone{chisqr.pdf}
%\caption{The reduced-$\chi^2$ as a function of eccentricity for the orbit of $\pi$~Men~d.}
 %\label{fig:pimen-d_chisqr}
%\end{figure}

%\begin{figure}[ht!]
%\plotone{planetdorbits.pdf}
%\caption{Orbital phase plots and fits  of the $\pi$ Men d for the best fit solution  $e$ = 0.4 (top),
%$e$ = 0.2 (middle) and $e$ = 0.0 (bottom).}
 %\label{fig:pimen-d_phase}
%\end{figure}

\section{The K-amplitude of $\pi$\,Men\,c}

Having determined the presence of all stable periodic signals in the RV time series we now focus on measuring a precise mass for the transiting planet $\pi$\,Men\,c. This hinges on the radial velocity K-amplitude, $K_\mathrm{c}$.

\newpage

\subsection{Pre-whitening and Joint Fitting}

\begin{figure}[ht!]
\plotone{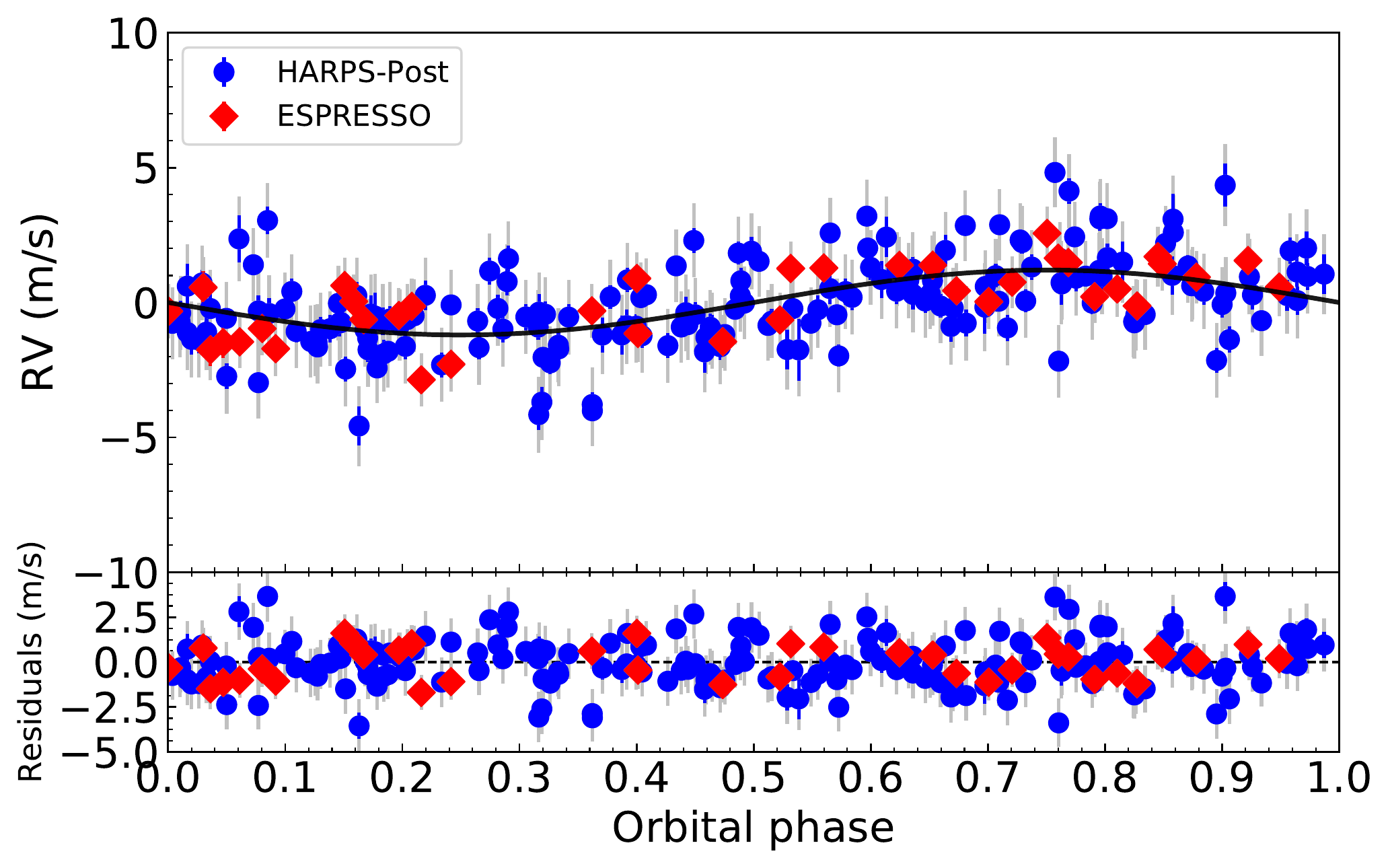}
\caption{Phase-folded RV variations and Keplerian orbit for $\pi$\,Men\,c. The contributions from planets b and d have been removed. Blue circles and red diamonds are nightly binned HARPS-POST and ESPRESSO RV measurements, respectively. The  gray  error bars include  jitter. The curve represents the orbital solution. The zero phase of the RV curve corresponds to the time of first transit (see Table\,\ref{tab:PiMen_d}).
 \label{fig:PiMen_c}}
\end{figure}

The simplest procedure is to fit sequentially a Keplerian orbit to each periodic signal, remove it, and fit the next periodic signal in a so-called pre-whitening procedure. We did this using only the HARPS-POST and ESPRESSO RVs, as these have the lowest rms scatter in our data-set. %We first fitted the orbit of the outer planet, removed this and then fitted the orbit of the transiting planet. A fit was then made for planet d on the residual RVs. The orbits of planets b and d was then subtracted from the original data and a fit to the orbit of planet c performed on the residual data. 
We first fitted the orbit of the outer planet ($\pi$\,Men\,b), removed this and then fitted the orbit of the planet at 125-d ($\pi$\,Men\,d). A fit was then made for planet c on the residual RVs. This procedure results in $K_\mathrm{c}$\,=\,1.30\,$\pm$\,0.13~\ms.
%1.26 0.14
An improved approach is to perform a joint fit an all Keplerian orbits that are present. This results in   $K_\mathrm{c}$\,=\,1.21\,$\pm$\,0.12~\ms, a value  in agreement with the result from the pre-whitening procedure to well within the nominal uncertainty. We adopt this as our best solution. Figure~\ref{fig:PiMen_c} shows the phase-folded Doppler reflex motion induced by $\pi$~Men~c on the star, following the subtraction of the RV signals of the other two planets.

\subsection{Floating Chunk Offset Method}

As an independent approach to determining the $K$-amplitude we applied the so-called floating chunk offset (FCO) method. This technique was developed in order to extract the Doppler reflex motion induced by ultra-short period (USP) planets, i.e., planets with orbital periods shorter than one day \citep{2010A&A...520A..93H}. The basis is that if one takes several measurements in one night the dominant variations come from the stellar orbital motion induced by the short-period planet. If the RV variations from other sources (rotation, long period planets, systematic errors, etc.) are much longer than the short-period planet, then the nightly variations stem predominantly from orbital motion of the USP planet and all other phenomena merely add a constant value to the RV for that night. One fits a Keplerian orbit keeping the period and phased fixed, but varying the $K$-amplitude and the zero-point offset velocity for each night (i.e. chunk). 

The FCO method can also be applied to planets with orbital periods longer than one day. The only criteria are that period of the transiting planet be smaller than periods from other sources and the observing cadence is reasonably high. This is the case for $\pi$\,Men\,c whose orbital period is much less than the 2088-d and 125-d periods present in the RV data. Although our inferred rotation period of 18 d and its harmonics could have an influence, there is no evidence for these in the periodogram of the RV time series. Our observing strategy for $\pi$~Men was such that RV measurements were taken on several consecutive nights. The advantage of the FCO method is that we do not have to remove the large orbital motion of the outer planet as it is naturally filtered out in the analysis.

\begin{figure}[ht!]
\plotone{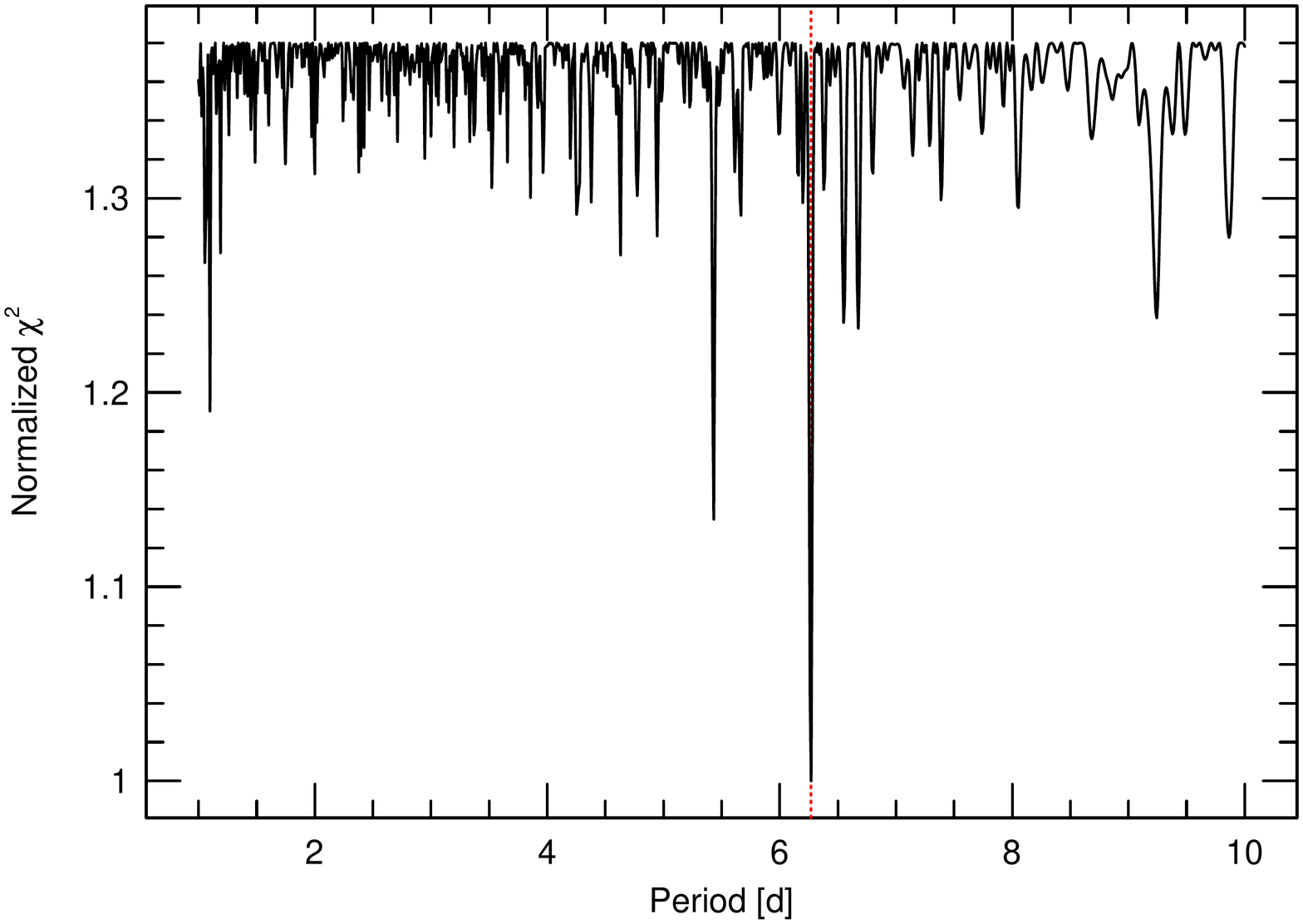}
\caption{The FCO periodogram of the raw RV data. The red dashed vertical line marks the orbital period of the transiting planet.}
 \label{fig:fco_periodogram}
\end{figure}

We divided the RV data from the HARPS large program into time chunks with consecutive measurements spanning two to four days. Figure~\ref{fig:fco_periodogram} shows the FCO periodogram   produced by fitting the RV chunk data using a trial period,  finding the best amplitude and plotting the reduced $\chi^2$ as a function of input period. The best fit was obtained for the period of the transiting planet; the FCO method can detect the signal of the $\pi$~Men\,c. We note that the raw RV data with the large orbital motion of the outer planet as well as the respective instrumental offsets were present in the chunks, so it is relatively insensitive to the details in how the other signals are removed. This resulted in $K_\mathrm{c}$\,=\,1.29\,$\pm$\,0.22~\ms, a value consistent with the results from other methods. 

A more refined value to the FCO result can be obtained by first subtracting the large orbital RV variations due to the outer planet ($\pi$\,Men\,b). The high amplitude and eccentricity can still result in a large, short-term variation, especially since our measurements caught the maximum in the orbital curve where the RV changes by many \ms\ over several days. This can introduce a systematic error in the $K$-amplitude determination. To remove the variations of $\pi$\,Men\,b we simply subtracted the orbital solution from Table~\ref{tab:orbits} from the raw data, which retained all the zero point offsets of the individual data sets. For this final step we  included the ESPRESSO measurements in the analysis. This resulted in $K_\mathrm{c}$\,=\,1.16\,$\pm$\,0.13\,\ms.

%\begin{figure}[ht!]
%\plotone{fcoampin.pdf}
%\caption{Input $K$-amplitude versus the output $K$-amplitude for $\pi$ Men c using the FCO method. The error bars
%represent the standard deviation of 50 runs using different random numbers.}
% \label{fig:fcoamp}
%\end{figure}

As a test to ensure that FCO recovers known input signals we created synthetic simulated data consisting of the Keplerian orbits  of planets c and d. These were sampled and divided into chunks in the same manner as the data. The input $K$-amplitude of $\pi$\,Men\,c was varied between 0 -- 5~\ms. The appropriate random noise was added, and to test an extreme case,  additional random offsets of several km\,s$^{-1}$ were added to  each chunk. Over the full range of the considered $K$-amplitudes the FCO method was able to recover the input amplitude.

One advantage of the FCO method is that, unlike for other methods, it can produce good results on sparse data where you do not have a good knowledge of timescales of other signals in the data. As a test, we took a subset of only 27 measurements taken over a time span of 3400~d and applied FCO. It yielded $K_\mathrm{c}$\,=\,1.40\,$\pm$\,0.49\,\ms, a $\sim$3\,$\sigma$ result that is consistent to within the error of our nominal value. Table~\ref{tab:k-amplitude} lists the $K_\mathrm{c}$-amplitude determinations using the different methods (pre-whitening, joint fit, and FCO). %These have a mean value of 1.21\,$\pm$\,0.05~\ms, which indicates that $K_\mathrm{c}$ amplitude is robust and insensitive to the method one uses to extract it.
These have a mean value of 1.22~\ms\ and agree well within their nominal uncertainties, which indicates that the $K_\mathrm{c}$ amplitude is robust and insensitive to the method one uses to extract it.

\subsection{K-amplitude from Individual Data Sets}

It is instructive to compare the $K$-amplitude of the transiting planet $\pi$\,Men\,c derived from individual and combinations of the data sets. In particular, single instrument sets will have the same zero point offset and systematic errors. We can see how consistent these solutions and their errors are to the nominal values.

For this exercise we first used the complete data set to fit and remove the orbital motion of the two outer planets. This was done because each data sets sample different parts and a smaller fraction of the longer period orbits and we want a fair comparison. Using subsets of the data to fit all planetary orbits will have a strong influence on $K_\mathrm{c}$. The remaining residuals contained just the orbital variations of the inner planet.

Table~\ref{tab:K_c-compare} shows the resulting $K_\mathrm{c}$ amplitudes for the individual data sets. All have consistent values with each other and to our nominal value. The HARPS-PRE data, however, give a larger amplitude and error compared to the ESPRESSO result in spite of having a comparable number of measurements. It is not known whether this is due to a higher intrinsic stellar jitter during this time, or to the different quality of the optical fiber used for scrambling. Note that the ESPRESSO amplitude of $K_\mathrm{c}$\,=\,1.25\,$\pm$\,0.24~\ms\ is slightly lower than the value of 1.5\,$\pm$\,0.2~\ms\ found by \citet{Damasso2020}. This may be due to authors not removing the underlying 125-d signal, which could not be seen in their data due to insufficient sampling. They did note the possible presence of a signal at $\approx$\,194\,d, but this was not significant and they chose not to include it in the modeling.

Adding the ESPRESSO data to the HARPS-POST data does not alter the $K$-amplitude significantly. In spite of the larger scatter of the HARPS-PRE data these also have little influence on the result when combining all the data sets. Although not listed in the table, we also performed a solution including the lower quality UCLES data and this resulted in $K_\mathrm{c}$\,=\,1.24\,$\pm$\,0.12~ \ms. Clearly, the solution is driven by the higher quality HARPS and ESPRESSO measurements. 

It is of interest to make a direct comparison between the performance of HARPS and ESPRESSO for comparable data taken at roughly the same time. To do this we took a subset of our HARPS measurements so that they had the same number as those from ESPRESSO (37  nightly averages) and roughly covered the same time span. This resulted in a $K_\mathrm{c}$\,=\,1.16\,$\pm$\,0.34~\ms, which compares favorably to our value of 1.25\,$\pm$\,0.24~\ms\ using the ESPRESSO data. For bright targets where the $S/N$ ratio is high, HARPS should perform as well as ESPRESSO for RV measurements.

\subsection{The mass of $\pi$\,Men\,c}

The $K_\mathrm{c}$-amplitude from the various methods and data sets (Tables~\ref{tab:k-amplitude} and \ref{tab:K_c-compare}, excluding the UCLES and HARPS-PRE RV measurements) %are all tightly clustered around a mean value of 1.21\,$\pm$\,0.03~\ms\ 
are all consistent and tightly clustered around a mean value of 1.21~\ms, indicating a well constrained $K_\mathrm{c}$-amplitude and thus planet mass. Since all the RV measurements were taken over a relatively short time span with excellent sampling we take the   $K_\mathrm{c}$-amplitude determined using a joint fit to the HARPS-Post + ESPRESSO data sets,  namely $K_\mathrm{c}$\,=\,1.21\,$\pm$\,0.12\,m\,s$^{-1}$, as our adopted value. Using the stellar mass of $M_{\star}$\,=\,1.07\,$\pm$\,0.04\,$M_\odot$ from \citet{Damasso2020} results in a planet mass, $M_\mathrm{c}$\,=\,3.63\,$\pm$\,0.38~M$_\oplus$. 

\citet{Damasso2020} derived a ratio of the planet to star radii of $R_\mathrm{c}/R_\star$\,=\,0.0165\,$\pm$\,0.0001 and a stellar radius of $R_\star$\,=\,1.17\,$\pm$\,0.02~R$_\odot$. Recently Csizmadia et al. (2021; submitted) showed that the color indices of a star can be used to check stellar parameters. For $\pi$\,Men they determined a stellar radius of   $R_{\star}$\,=\,1.190\,$\pm$\,0.004~R$_\odot$. This results in a planet radius of $R_\mathrm{c}$\,=\,2.145\,$\pm$\,0.015~R$_{\oplus}$ and a bulk density of $\rho_\mathrm{c}$\,=\,2.03\,$\pm$\,0.22~g\,cm$^{-3}$ for $\pi$\,Men\,c.

We note that \citet{2018ApJ...868L..39H} and \citet{2018A&A...619L..10G} derived a planet radius of $R_c$ = 2.21 $\pm$ 0.04 R$_\oplus$ and $R_c$ = 2.23 $\pm$ 0.04 R$_\oplus$, respectively. This results in a lower planet density of $\rho_c$ = 1.8 $\pm$ 0.2~g\,cm$^{-3}$. This only highlights that even though we have formal error of 0.7 \% in the planet radius, the true uncertainty may be larger.

\begin{figure}[ht!]
\plotone{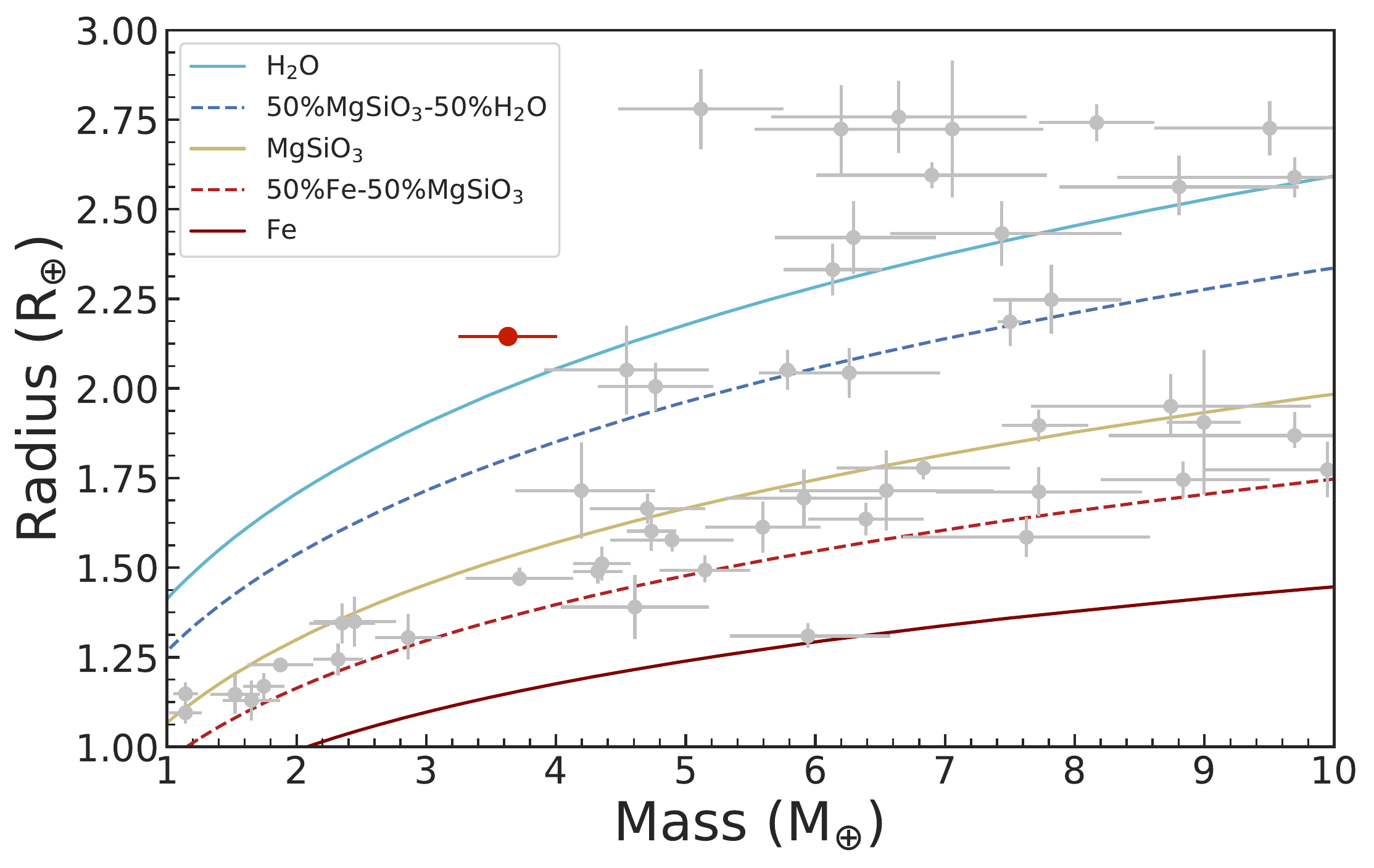}
\caption{The location of $\pi$ Men c in the mass-radius diagram for exoplanets with well determined radii and masses (better to 15\,\%). Composition models from \citet{Zeng2016} are displayed with different lines and colors.}
 \label{fig:M-R}
\end{figure}

\begin{figure}[ht!]
\plotone{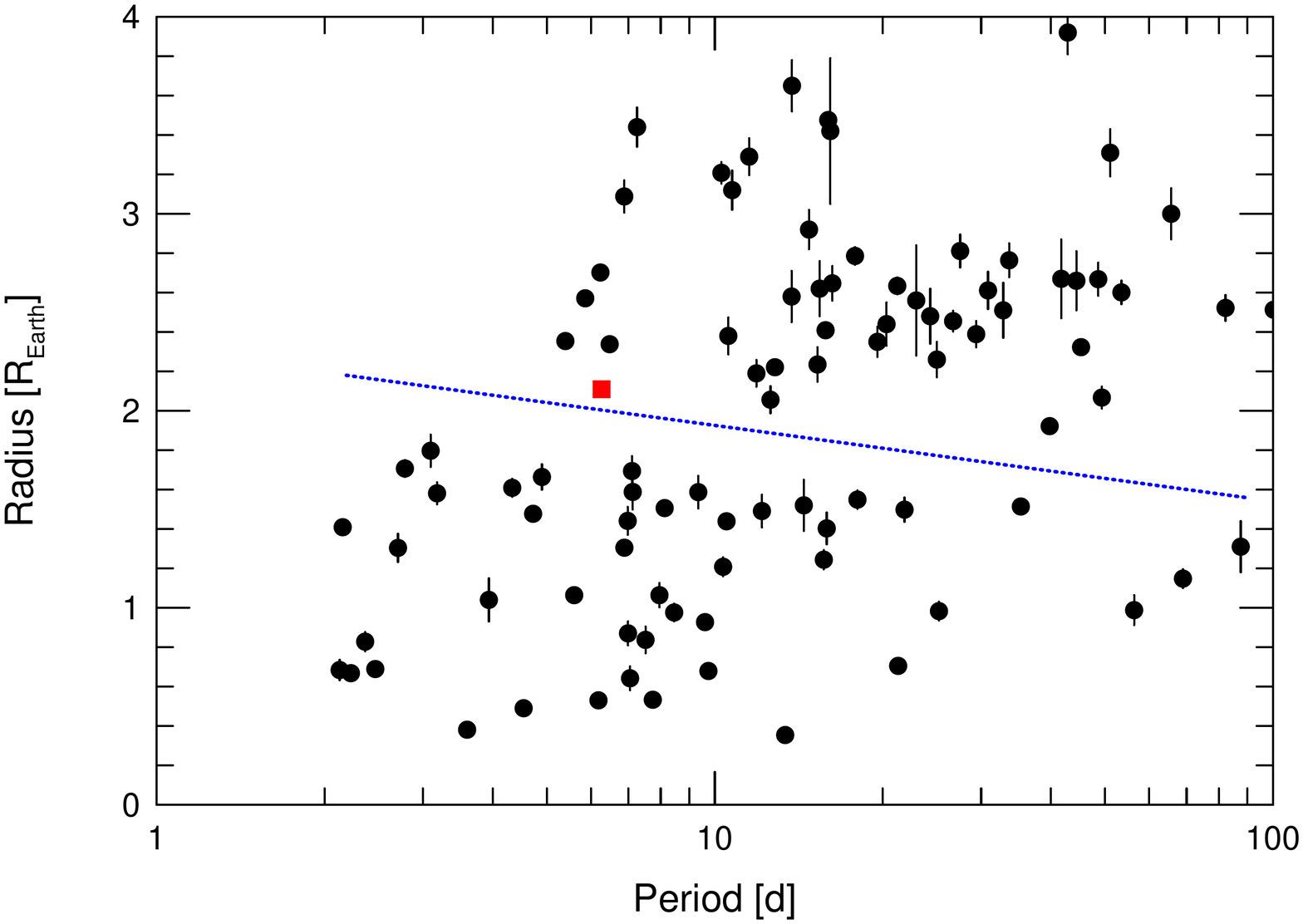}
\caption{The location of the radius valley from \citet{2018MNRAS.479.4786V}. The blue dashed line marks the hyperplane of maximum separation. The location of $\pi$\,Men\,c is marked by the red  square.}
\label{fig:RV}
\end{figure}

\section{Discussion}

Our investigation of the RV variations of $\pi$ Men has produced two main results - an accurate and precise mass for the transiting planet $\pi$ Men\,c and the discovery of a third planet in the system.

\subsection{$\pi$\,Men\,c in the Mass-Radius Diagram }

With a bulk density of 2.03\,$\pm$\,0.22~g\,cm$^{-3}$, $\pi$ Men\,c appears to have a rather low density compared to most exoplanets with its mass. This is highlighted by Figure~\ref{fig:M-R} which shows the location of $\pi$\,Men\,c in the mass-radius diagram for exoplanets with well determined radii and masses (better to 15 \%). $\pi$\,Men\,c lies closest to the internal structure model for pure water. We stress that these are simple planet composition models for a planet without an atmosphere and, as noted \citet{2014ApJ...792....1L}, an atmospheric envelope may only account for a few percent of the planet mass, but it can have a huge impact on the measured planet radius. 

{With its relatively large radius $\pi$~Men~c means it probably held a significant volatile envelope and thus is a prime target for atmospheric
characterization \citep{2021arXiv210809109H}. \citet{2020ApJ...888L..21G} searched for hydrogen from photodissociation in the atmosphere of
$pi$~Men~c using Ly-$\alpha$ in-transit spectroscopy with the {\it Hubble Space Telescope} (HST)  but found none. One explanation for the non-detection was that the atmosphere was dominated by
water or other heavy molecules rather than H$_2$/He. Subsequently, \citet{2021ApJ...907L..36G}, again using {\it HST},  were able to
detect C II ions at the 3.6-$\sigma$ level which seems
to support this scenario. The planet may still be in transition into a bare rocky planet. More progress on the characterization of
the atmosphere of $\pi$-Men~c will surely come with investigations using the successfuly launched  {\it James Webb Space Telescope}. }

{The planet lies near the middle of the so-called radius valley,   a gap in the radius distribution of small planets around 1.75\,-\,2.00~R$_\oplus$ that
 separates planets with masses sufficient to maintain a H-He envelope from those  without such an envelope.  The orbital period of $\pi$\,Men\,c and its radius places it on the boundary between the two classes of planets (Fig.~\ref{fig:RV}) It is still an open
question as to whether this gap results from planet formation or evolution. One hypothesis is that it is
 caused by atmospheric erosion of short-period planets due to photoevaporation  from the close proximity to the host star (e.g. \citealt{2017ApJ...847...29O}; 
 \citealt{2017AJ....154..109F} \citealt{2018AJ....156..264F}). Alternatively, the gap may simply result from the planet formation process
 itself.  \cite{2018MNRAS.476..759G} showed that planet formation with a core-powered mass loss mechanism %\citep{2016ApJ...825...29G}  
 could account for the radius distribution of planets without invoking photoevaporation. This 
 scenario is able to match the radius valley, location, shape and slope \citep{2019MNRAS.487...24G}.
 
Clearly, the addition of just a single point in the period-radius diagram will not provide the breakthrough in understanding the origin of radius valley. However, exoplanets with
precisely determined masses and radii, in particular those in the middle of the gap like $\pi$~Men~c, are needed shed more light on the cause of the gap.}

\subsection{$\pi$\,Men\,d}

Our analysis of the RV time series reveals the presence of a 125-d periodic signal in the data very likely due to a third planet in the system. Orbital solutions yield a minimum planet mass of $M_\mathrm{d}$\,sin\,$i_\mathrm{d}$\,=\,13.38\,$\pm$\,1.35~M$_\oplus$ in an eccentric orbit (e\,=\,0.220\,$\pm$\,0.079). A preliminary dynamical study indicates that its orbit is stable on time scales up to 20\,Myrs, {at least over the orbital parameters that we have probed.}  In order to be certain this RV signal comes from a planet it must satisfy three criteria: 1) The signal should be statistical significant with a FAP\,$<$\,0.1\%. 2) The signal should be long-lived and coherent with no change in the period, amplitude and phase. 3) There should be no variations with the RV with any indicators (photometry, Ca II, etc.), which would suggest a stellar origin for the variations. It appears that the 125-d signal satisfies these criteria.

The signal is highly significant with a FAP\,$\ll$\,3.3\,$\times$\,10$^{-6}$, which makes it unlikely to arise from noise. It also appears to be stable and coherent. { Adding measurements causes the statistical significance as shown  by the $P$-$N$, to  increases in the expected manner given the signal, sampling, and noise level. } There may be a slight concern that the slope of the $P$-$N$ behavior is shallower than expected from simulations, but this may be due to the noise not following a strictly a Gaussian distribution. We stress, however, that such a stable growth  in $P$-$N$ should not be used as a final  confirmation of the planetary nature of an RV signal. A case in point is the proposed planet around $\alpha$\,Tau. \citet{2015A&A...580A..31H} found evidence for a 629-d RV signal whose power behavior in the $P$-$N$ diagran followed the expected growth for a stable signal. In spite of this, additional RV measurements seemed to contradict  the planet hypothesis \citep{2019A&A...625A..22R}.

There seems to be no clear variations with the RV period in the activity indicators. The S-index shows periodic variations with a period ($P$\,$\sim$\,500\,d) which is much longer than RV value. The FWHM shows a dominant peak in the periodogram at $\sim$200~d, while the BIS shows a period of $\sim$ 760 d. There is a weak feature in the periodogram of the BIS measurements is coincident with the RV signal, but as discussed earlier, we do not deem this as significant.

The subset of S-index variations as well as the inferred value from the radius and rotational velocity of the star indicate a stellar rotational period of approximately 18 d. Clearly, the 125-d is not due to rotational modulation. If an activity cycle is present it most likely has a period of 500-700 d, as manifested in the activity indicators. G-type stars are expected to have activity cycles with timescales of  years to decades. However, shorter period activity cycles are not unprecedented for other late-type stars. \citet{2017A&A...600A.120S} found evidence for a $\sim$120-d cycle in the F6 star $\tau$ Boo. One may speculate that the 125-d RV period is roughly one-fourth of an activity cycle of $\sim$ 500 d and we cannot exclude this with certainty. However, it is puzzling that the third harmonic\footnote{Analogous to stellar oscillations, we refer to the rotation  period, $P$, as the ``fundamental'' and $P/4$ as the third harmonic.} dominates the RV, yet the ``fundamental'' period dominates the S-index measurements. 

We should stress that although the 125-d signal seems to satisfy our criteria for planet confirmation, all of these criteria are {\it necessary}, but not {\it sufficient} conditions for planet confirmation. That is to say that an RV planet candidate must satisfy these criteria, but that is still no guarantee that the signal stems from the orbital motion of a planet. This is especially true for weak signals with RV amplitudes of  a few $\sim$\,m\,s$^{-1}$. Conceivably, a stellar activity cycle or its harmonics could be seen in the RV data without a strong presence in classic activity indicators. We are just starting to understand the influence of stellar activity on precise RV measurements and the time scales involved, so surprises could be in store. However, all the best available evidence at hand suggests that  the presence of a third planet in the system is the most likely explanation for the 125-d signal in the RV data. 

\begin{figure}[ht!]
\plotone{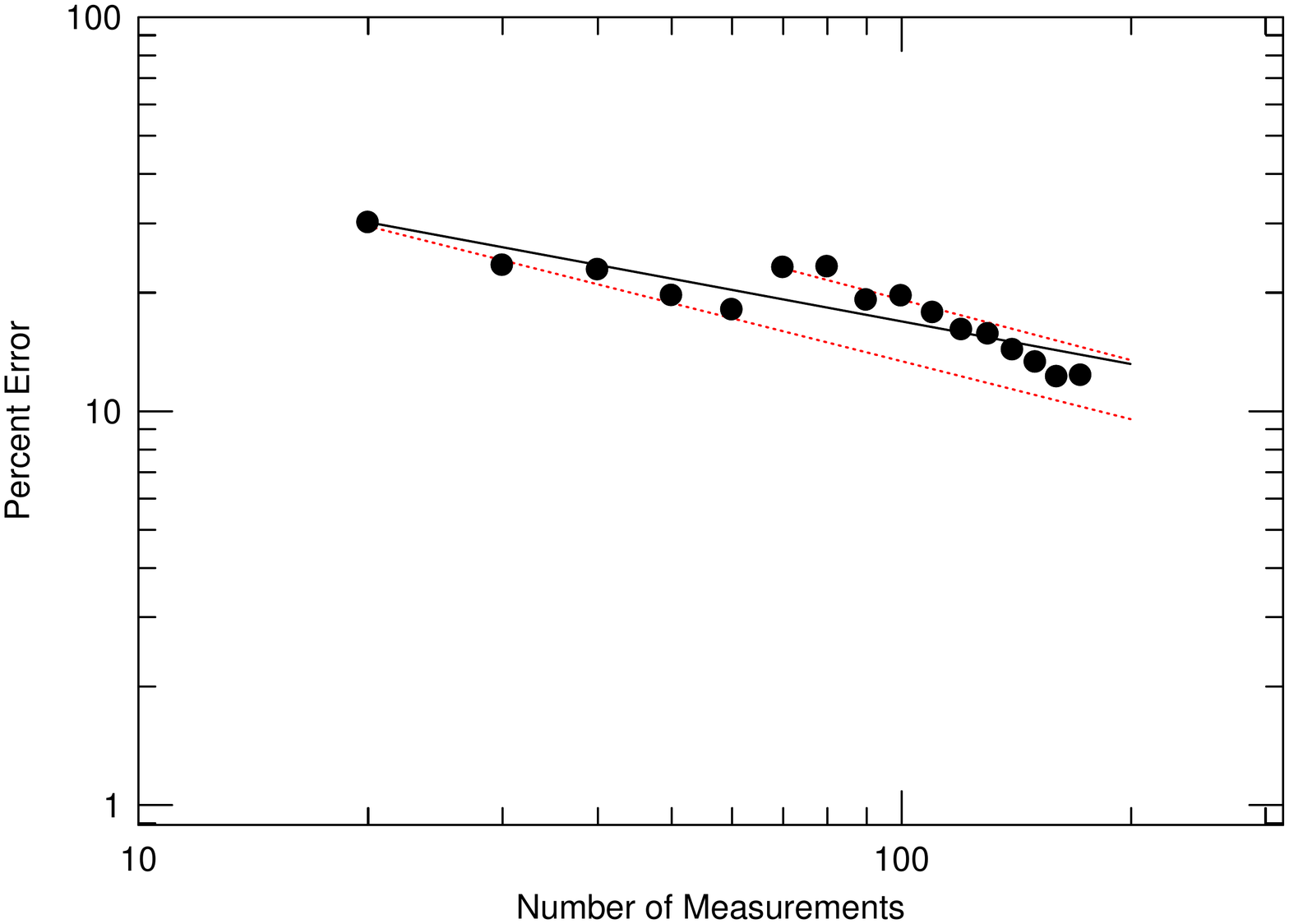}
\caption{The error in the mass determination as a function of the number of measurements. The solid line represents a fit with error $\propto\,N^{-0.36}$. The red dashed lines represent a nominal error $\propto\,N^{-0.50}$ expected for white noise. The top red dashed line has its origin at the $N$\,=\,70 data point.}
\label{fig:err_v_n}
\end{figure}

{The planet $\pi$\,Men\,d  may have implications  for the formation of the innner planet. \citet{2020A&A...640A..73D} argued that the presence of nearby planets to $\pi$\,Men\,c would favor in-situ formation. At first glance, it would seem difficult  for $\pi$\,Men\,c to have  formed in the outer region of the proto-planetary disk and then migrated inwards if other planets were present. However, one could envision a scenario where $\pi$~Men~c formed first, migrated inwards and then $\pi$~Men~d formed at a later time. Dissipation  of the disk would then halt the migration of $\pi$~Men~d.

Alternatively, the inner planet could have formed via tidal migration. An interaction of two planets (say c and d) would have scattered planet c into the inner regions
where its orbit would have been circularized via tidal effects. The planet d remained, but in a highly eccentric orbit. The outermost giant planet
can also perturb the inner planets. Various effects of the giant planet
on $\pi$~Men~c were investigated by \cite{2020MNRAS.497.2096X} and
\cite{2020A&A...640A..73D}. 
However, here is not the place to speculate on specific formation scenerios. That is best left for detailed theoretical modeling which can produce the observed configuration
of the $\pi$~Men planetary system. To do that requires knowing the full architecture of the planetary system and $\pi$~Men~b is a key component.  It is therefore important to derive more accurate orbital elements for $\pi$\,Men\,d. }

Our simple dynamical analysis of the planetary system to $\pi$ Men indicates that the orbit of planet d should be stable. Clearly, a more detailed dynamical analysis is required, which is best left for a dedicated investigation. The $\pi$~Men planetary system offers us a very interesting system for such studies, especially given the relative high mass of the outer planet and its inclined orbit. 
%It is of interest to see if the outer planet can induce Kozai oscillations \citep{1962AJ.....67..591K} in the inclination and eccentricity of planet d.

\subsection{Lessons learned for the RV follow-up of small transiting planets}

As a closing remark, our study of $\pi$\,Men offers us important lessons for the follow-up of small transiting planets in the era of TESS, and in the near future, PLATO. First, $\pi$\,Men\,c has one of the most precise mass determinations for a small planet, with an error of about 10\,\%. This precision required approximately 200 (nightly averaged) measurements taken with superb instruments on a bright star. If one wants to increase the precision on the mass measurement one naturally requires more measurements, but the number of these may not always follow the expectations from white noise.

Figure~\ref{fig:err_v_n} shows the percent error in the mass of $\pi$\,Men\,c as a function of the number of RV measurements, $N$, using the homogeneous data set provided by the HARPS-Large RVs. For white noise one expects an error proportional to $N^{-0.5}$, but in reality it is slightly worse, being $\propto$ $N^{-0.36}$. At the beginning our RV measurements performed   as expected, but at about $N$\,=\,70 the error in the mass measurement increased. For subsequent measurements, the error does follow the expected behavior $\sigma$ $\propto$ $N^{-0.5}$, but from a higher starting point. There must be a ``red'' noise component, most likely attributable to a variable contribution of stellar jitter. One cannot always rely on a few more RV measurements to improve the error on the $K$-amplitude. 

In the case of  $\pi$\,Men\,c, improving the mass measurement to better than 5\,\%  would require more than 1000 RV measurements if one were to adhere to the same observing strategy and sampling of our study. This is 50 \% more than the number estimated based on the trend shown at the start of the measurements. Clearly, if one is only interested in deriving the mass of the transiting planet it is more effective to obtain a large number of RV measurements for $\pi$\,Men over a much short time span, preferably over a single orbital period. Even then, a considerable number of observations will be required to find and remove additional signals due to stellar rotation or other planets which also can have a large influence on the $K$-amplitude of the transiting planet. This is demonstrated by the ESPRESSO data. \cite{Damasso2020} found a $K_c$-amplitude 20\% lower than our value for the same data set, most likely by not including
the Keplerian motion of the third planet. Considerably more RV measurements were required to be certain of the presence of the 125-d signal. It is clear that  the precise mass determinations of small planets will come with at a very steep price even for bright targets.

Second, $\pi$\,Men seems to be a relatively
inactive star, at least at the time scales of the transiting planet's orbital period. It is also a bright star which results in high signal-to-noise ratio data acquired in short exposure times. In spite of this, the resulting rms scatter for the RV measurements is $\approx$1.5\,\ms\ even after removing all periodic signals. This is a factor of 2-3 larger than the measurement errors, even when using the {premier instruments  in the world for RV measurements.} Regardless of the effort taken to minimize instrumental and photon noise, the noise floor will be set by the intrinsic variability of the star. We will be hard pressed to find stars that are ``RV quieter'' than about 1\,\ms\ and most stars will have an intrinsic jitter higher than this.

For bright targets with high stellar jitter, telescope size will not necessarily matter. The RV confirmation of small  transiting planets will require an inordinate amount of telescope resources. $\pi$\,Men has shown us that for bright targets with high intrinsic jitter, using a larger aperture telescope does not always result in higher precision measurements. PLATO is expected to produce a large number of small transiting planet which will severely stress the available telescope resources.  Instruments on 2-3m class telescopes that  provide an RV measurement precision of 3-5 m\,s$^{-1}$ can clearly play an important role in the PLATO era. The lack in measurement precision can be compensated with the shear number of measurements that  can be invested on one target. A simple simulation shows that with $\sim$150 RV measurements over nine consecutive nights one can recover the $K$-amplitude of $\pi$\,Men\,c (4$\sigma$ result) with a modest measurement precision of 3~\ms. Bringing more 2-3m class telescopes to the upcoming PLATO follow-up effort could play an important role in the success of the mission.

\begin{table}
\centering
\footnotesize
\caption{Equatorial coordinates, proper motion, parallax, distance, $V$-band magnitude, and fundamental parameters of $\pi$\,Men.}
\label{tab:star_parameters}
%\centering
\begin{tabular}{lrr}
\hline\hline
\noalign{\smallskip}
Parameter & Value & Source \\
\noalign{\smallskip}
\hline
\noalign{\smallskip}
\multicolumn{3}{l}{\emph{Equatorial coordinates and $V$-band magnitude}} \\
\noalign{\smallskip}
RA (hh:mm:ss, J2000) & 05:37:09.885 & \citet{GaiaDR2} \\
RA (dd:mm:ss, J2000) & -80:28:08.831 & \citet{GaiaDR2} \\
$V$ & $5.65 \pm 0.01$ & \citet{Mermilliod1987}\\
\noalign{\smallskip}
\hline
\noalign{\smallskip}
\multicolumn{3}{l}{\emph{Proper Motion, Parallax, and Distance}} \\
$\mu_{\alpha} \cos \delta$ (mas \ yr$^{-1}$) & $311.187 \pm 0.127$ & \citet{GaiaDR2} \\
$\mu_{\delta} $ (mas \ yr$^{-1}$) & $1048.845 \pm 0.136$ & \citet{GaiaDR2} \\
Parallax (mas) & $ 54.7052\pm0.0671$ & \citet{GaiaDR2} \\ 
Distance (pc) & $18.280\pm0.022$& \citet{GaiaDR2} \\ 
\noalign{\smallskip}
\hline
\noalign{\smallskip}
\multicolumn{3}{l}{\emph{Fundamental parameters}} \\
Star mass $M_{\star}$ (M$_\odot$) & $1.07 \pm 0.04$  & \citet{Damasso2020} \\
Star radius $R_{\star}$ (R$_\odot$) &  $1.190 \pm 0.004$ & 
Csizmadia et al. (2021, submitted) \\
Effective Temperature $\mathrm{T_{eff}}$ (K) & $5998 \pm 62$ & \citet{Damasso2020} \\
Iron abundance [Fe/H] (dex)                   & $0.09\,\pm\,0.04$ & \citet{Damasso2020} \\
Projected rotational velocity \vsini\ (\kms)  & $3.34\,\pm\,0.07$ &  \citet{Damasso2020} \\
Age (Gyr)                                     & $3.92^{+1.03}_{-0.98}$ & \citet{Damasso2020} \\
\noalign{\smallskip}
\hline
\end{tabular}
%\tablecomments{(a) \gaia\ Data Release 2 \citep[DR2;][]{GaiaDR2}.}
\end{table}

\begin{deluxetable}{lccc}
%\tablenum{2}
\tablecaption{The RV Data Sets.}
\tablewidth{0pt}
\tablehead{
\colhead{Data Set} &  \colhead{Time Span (Year) }  & \colhead{ Measurements (nightly)} & \colhead{$\sigma$ (m\,s$^{-1}$)} 
 }
\startdata
UCLES & 1988.04 -- 2015.9 & 77 & 5.88  \\
HARPS-PRE   & 2004.0  -- 2015.0 & 31 & 2.72 \\
HARPS-POST (Archival) & 2015.8 -- 2016.2 & 9 &  0.83  \\
HARPS-POST (Large) & 2018.7 -- 2020.2 & 177 & 1.40 \\
ESPRESSO  & 2018.7 -- 2019.2   &  37 & 0.95 \\
\enddata
%\tablecomments{}
\end{deluxetable}
\label{tab:rv_sets}

\begin{deluxetable}{lrrrrrr}
%\tablenum{2}
\tablecaption{RV measurements and activity indicators from  the ESO HARPS Large Programs (full table in electronic version)}
\tablewidth{0pt}
\tablehead{
 \colhead{BJD$_\mathrm{TDB}$}  & \colhead{RV} & \colhead{$\sigma_\mathrm{RV}$} & \colhead{BIS} & \colhead{FWHM} & \colhead{S-index} & \colhead{$\sigma_\mathrm{S-index}$}  \\
 \colhead{(d)} & \colhead{\ms} & \colhead{\ms} & \colhead{\ms} & \colhead{\kms} & & }
\startdata
2458383.896397 &  10994.6 &  0.4 &  8.974 &  7.6796 &  0.153485 &  0.000362 \\
2458383.899499 &  10994.5 &  0.4 &  9.763 &  7.6807 &  0.153990 &  0.000368 \\
2458384.814003 &  10998.8 &  0.4 &  9.695 &  7.6768 &  0.153351 &  0.000338 \\
2458384.817116 &  11000.3 &  0.4 &  8.031 &  7.6791 &  0.153506 &  0.000362 \\
2458385.813365 &  11003.3 &  0.4 &  8.265 &  7.6798 &  0.153090 &  0.000287 \\
2458385.816513 &  11003.4 &  0.4 &  8.884 &  7.6788 &  0.153815 &  0.000288 \\
2458385.898502 &  11003.7 &  0.3 &  8.891 &  7.6788 &  0.152502 &  0.000327 \\
\enddata 
%\tablecomments{}
\end{deluxetable}
\label{tab:rv_data}

\begin{deluxetable*}{l|c|c}
%\tablenum{1}
\tablecaption{Orbital Parameters for $\pi$\,Men\,b and c.}
\tablewidth{0pt}
\tablehead{
\colhead{Parameter} & \colhead{$\pi$ Men b} & \colhead{$\pi$ Men c} }
%%\decimalcolnumbers
\startdata
Orbital period $P_\mathrm{orb}$ [d]                            & 2088.33 $\pm$ 0.34                                &  6.267852 $\pm$ 0.000016 \\ 
Time of inf. conj. / first transit $ T_0$ [BJD$_\mathrm{TDB}$ - 2450000 d]        & 6540.34   $\pm$ 0.75                        & 8519.8068  $\pm$ 0.0003   \\
Orbit eccentricity $e$                                      & 0.6396 $\pm$ 0.0009                                  & 0 (fixed)                        \\
Argument of periastron of stellar orbit $\omega_\star$  [deg]           & 331.03 $\pm$ 0.25                                 & 90 (fixed)                    \\
Radial velocity semi-amplitude variation $K$  [\ms]                & 192.99 $\pm$ 0.38                                 & 1.21 $\pm$ 0.12        \\
%Mass function [M$_\odot$]  & (6.97  $\pm$ 0.04) $\times$ 10$^{-7}$   & (1.18 $\pm$  0.38 ) $\times$ 10$^{-15}$  \\
Planetary minimum mass $M_\mathrm{p}$\,$\times$\,sin\,$i_\mathrm{p}$         & 9.82  $\pm$ 0.24 M$_\mathrm{Jup}$                  & 3.63 $\pm$ 0.38  M$_\oplus$\\
True$^1$ planetary mass $M_\mathrm{p}$                                     & 12.6  $\pm$ 2.0  M$_\mathrm{Jup}$                 & 3.63 $\pm$ 0.38  M$_\oplus$\\
\enddata
\tablecomments{~$^1$\,Using $i_\mathrm{b}$ = { 51.2$^\circ$\,$^{+14.1^\circ}_{-9.8^\circ}$}\ for $\pi$\,Men\,b \citep{2020MNRAS.497.2096X} and $i_\mathrm{c}$ = 87.05$^\circ$\,$\pm$\,0.15$^\circ$ for $\pi$\,Men\,c \citep{Damasso2020}.}
\end{deluxetable*}
\label{tab:orbits}

\begin{deluxetable}{lcc}
%\tablenum{2}
\tablecaption{Periods in Activity Indicators and Their False Alarm Probabilities.}
\tablewidth{0pt}
\tablehead{
\colhead{Indicator} &  \colhead{Period [d]}  & \colhead{FAP} 
 }
\startdata
S-Index   & 456\,$\pm$\,9      &  $<$\,5\,$\times$\,10$^{-6}$     \\
FWHM    &  184\,$\pm$\,2    &   $<$\,5\,$\times$\,10$^{-6}$ \\
BIS         &  757\,$\pm$\,64   & 0.0025 \\
\enddata
%\tablecomments{}
\end{deluxetable}
\label{tab:indicators}

\begin{deluxetable*}{ll}
%\tablenum{1}
\tablecaption{Orbital Parameters for $\pi$\,Men\,d.}
\tablewidth{0pt}
\tablehead{
\colhead{Parameter} & \colhead{Value} }
%%\decimalcolnumbers
\startdata
Orbital period $P_\mathrm{orb,d}$ [d] & 124.64$^{+0.48}_{-0.52}$   \\
Time of inferior conjunction $T_{0,\mathrm{d}}$  [BJD$_\mathrm{TDB}$ - 2450000 d]        & 7595.46$^{+6.90}_{-6.39}$  \\
Eccentricity $e_\mathrm{d}$ & 0.220 $\pm$ 0.079 \\
Argument of periastron of stellar orbit $\omega_{\star,\mathrm{d}}$  [deg] & 323$^{+25}_{-73}$ \\
Radial velocity semi-amplitude variation $K_\mathrm{d}$  [\ms]               & 1.68 $\pm$ 0.17 \\
%Mass function [M$_\odot$]   & (6.11 $\pm$  1.83) $\times$ 10$^{-14}$    \\
Planetary Minimum mass  $M_\mathrm{d}$\,$\times$\,sin\,$i_\mathrm{d}$ [M$_{\oplus}$]       & 13.38 $\pm$ 1.35      \\
\enddata
\end{deluxetable*}
\label{tab:PiMen_d}

%\begin{deluxetable}{lccc}
%\tablenum{2}
%\tablecaption{K-amplitudes in m/s  for Pi\,Men\,c}
%\tablewidth{0pt}
%tablehead{
%\colhead{Method} & \colhead{HARPS} & \colhead{ESPRESSO} & \colhead{Combined} 
%}
%\decimalcolnumbers
%\startdata
%Joint fit                   & 1.30  $\pm$ 0.11  & 1.23  $\pm$ 0.10 & 1.27 $\pm$ 0.10 \\
%Pre-whitening            & 1.18 $\pm$ 0.16  & 1.42 $\pm$ 0.20    & 1.26 $\pm$ 0.14 \\
%Simultaneous fitting  & ?                           & ?    & 1.25 $\pm$ 0.12  \\
%FCO	                         & 1.10 $\pm$ 0.15   & 1.31 $\pm$ 0.28     & 1.16 $\pm$ 0.13 \\
%FCO$_{corrected}$ & 1.25 $\pm$ 0.22   & 1.23 $\pm$ 0.20 & 1.34 $\pm$ 0.25 \\
%\enddata
%\tablecomments{}
%\end{deluxetable}
%$\label{tab:k-amplitude}

\begin{deluxetable}{lccc}
%\tablenum{2}
\tablecaption{K-amplitude for $\pi$\,Men\,c}
\tablewidth{0pt}
\tablehead{
\colhead{Method} & $K_\mathrm{c}$ [\ms]  
}
%\decimalcolnumbers
\startdata
Pre-whitening   &    1.30 $\pm$ 0.13 \\
Joint fit       &    1.21 $\pm$ 0.12  \\
FCO	            &    1.16 $\pm$ 0.13 \\
\enddata
%\tablecomments{}
\end{deluxetable}
\label{tab:k-amplitude}

\begin{deluxetable}{lcc}
\tablecaption{$K_\mathrm{c}$-Amplitude from  Data Sets}
\tablewidth{0pt}
\tablehead{
\colhead{Data Set} & \colhead{Measurements} & \colhead{$K_\mathrm{c}$ [\ms]}  
}
%\decimalcolnumbers
\startdata
HARPS-PRE   &   42    &  1.78 $\pm$ 0.60 \\
HARPS-Post  &   186   &  1.18 $\pm$ 0.15  \\
HARPS-Large &   177   &  1.20 $\pm$ 0.16 \\
ESPRESSO    &   37    &  1.25 $\pm$ 0.24 \\ 
\hline
HARPS-POST + ESPRESSO  & 223 & 1.20 $\pm$ 0.13 \\
HARPS-Large + ESPRESSO & 214 & 1.22 $\pm$ 0.13 \\
\enddata
%\tablecomments{}
\end{deluxetable}
\label{tab:K_c-compare}

\acknowledgments
This work was supported by the KESPRINT collaboration, an international consortium devoted to the characterization and research of exoplanets discovered with space-based missions (www.kesprint.science).SG, MP, SC, APH, KWFL, ME, and HR acknowledge support by DFG grants PA525/18-1, PA525/19-1, PA525/20-1, HA 3279/12-1 and RA 714/14-1 within the DFG Schwerpunkt SPP 1992, ``Exploring the Diversity of Extrasolar Planets". LMS and DG gratefully acknowledge financial support from the CRT foundation under Grant No. 2018.2323 ``Gaseous or rocky? Unveiling the nature of small worlds". JK  gratefully acknowledges the support of the Swedish National Space Agency
(SNSA; DNR 2020-00104). PGB acknowledges  the financial support by NAWI Graz. We thank Matthias Hoeft and Alexander Drabent for use of the computer of the Tautenburg radio group for our dynamical study.
\software{pyaneti (Barrag\'an et al. 2019), DRS (Lovis \& Pepe 2007), HARPS-TERRA
(Anglada-Escud\' e \& Butler 2012), Gaussfit (Jefferys et al. 1987)}

%\vspace{5mm}
%\facilities{HST(STIS), Swift(XRT and UVOT), AAVSO, CTIO:1.3m,
%CTIO:1.5m,CXO}

\bibliography{PiMen}{}
\bibliographystyle{aasjournal}

\end{document}